\documentclass[final,times,10pt]{elsarticle}

\usepackage{amsmath}
\usepackage{amsfonts}
\usepackage{amssymb}
\usepackage{geometry}
\usepackage{graphicx}
\usepackage[numbers]{natbib}
\usepackage{hyperref}                   

\begin{document}
\newcommand{\ud}{\mathrm{d}}
\newcommand{\uD}{\mathrm{D}}
\newcommand{\noun}[1]{\textsc{#1}}

\begin{frontmatter}

\title{FISH: A 3D parallel MHD code for astrophysical applications}

\author[auth1]{R. K\"appeli\corref{cor1}}
\cortext[cor1]{Corresponding author, Tel.: +41 61 267 3785, Fax: +41 61 267 3784.}
\ead{roger.kaeppeli@unibas.ch}
\author[auth1]{S. C. Whitehouse}
\author[auth1]{S. Scheidegger}
\author[auth2]{U.-L. Pen}
\author[auth1]{M. Liebend\"orfer}

\address[auth1]{
Physics Department,
University of Basel,
Klingelbergstrasse 82,
CH-4056 Basel,
Switzerland}
\address[auth2]{
Canadian Institute for Theoretical Astrophysics,
60 St. George Street,
Toronto, Canada
}

\begin{keyword}
Numerical methods \sep
Magnetohydrodynamics \sep
Constrained transport \sep
Operator splitting \sep
Relaxation schemes \sep
Source terms

\PACS
02.30.Jr \sep 
02.60.Cb \sep 
47.11.-j \sep 
47.65.-d \sep 
95.30.Qd \sep 
97.60.Bw 

\end{keyword}

\begin{abstract}
\noun{fish} is a fast and simple ideal magneto-hydrodynamics code that scales to $\sim 10\,000$
processes for a Cartesian computational domain of $\sim 1000^{3}$ cells. The simplicity
of\noun{fish} has been achieved by the rigorous application of the operator splitting
technique, while second order accuracy is maintained by the symmetric ordering of the
operators. Between directional sweeps, the three-dimensional data is rotated in memory so
that the sweep is always performed in a cache-efficient way along the direction of
contiguous memory. Hence, the code only requires a one-dimensional description of the
conservation equations to be solved. This approach also enable an elegant novel
parallelisation of the code that is based on persistent communications with MPI for cubic
domain decomposition on machines with distributed memory. This scheme is then combined
with an additional OpenMP parallelisation of different sweeps that can take advantage of
clusters of shared memory. We document the detailed implementation of a second order TVD
advection scheme based on flux reconstruction.
The magnetic fields are evolved by a constrained transport
scheme. We show that the subtraction of a simple estimate of the hydrostatic gradient
from the total gradients can significantly reduce the dissipation of the advection scheme
in simulations of gravitationally bound hydrostatic objects. Through its simplicity and
efficiency, \noun{fish} is as well-suited for hydrodynamics classes as for large-scale
astrophysical simulations on high-performance computer clusters. In preparation for
the release of a public version, we demonstrate the performance of \noun{fish} in a suite of
astrophysically orientated test cases.
\end{abstract}

\end{frontmatter}

\section{Introduction}
\label{sec:introduction}

It has been argued that computational models provide the third pillar
of scientific discovery, beside the traditional experiment and theory.
On the practical side, there are indeed a lot of similarities between
computer models and experiments: In both cases one tries to start
from a well-defined initial state. The outcome of the simulation is
in the same way unknown as the outcome of an experiment and unexpected
results are interesting in both cases. Also, one has to define observables
whose values one intends to measure at several time intervals or at
the end of the simulation or experiment. If the computer model or
experiment is complicated enough, these values will be affected by
systematic and statistical errors and their understanding requires
an involved theoretical analysis and interpretation. Finally, both
computer model and experiment need sufficient, and sometimes quite
extensive, documentation to make the results reproducible.

On the other hand, only the real experiment is capable of probing
the properties of nature, while the computer model is bound to elaborate
the properties of the theory it is built on. The strength of the computer
simulation is that it may demonstrate and extend the predictive power
of its underlying theory. In principle, though, it seems completely
irrelevant whether the consequences of the underlying theory are evaluated
by analytical approaches or a computer-aided model. The computer model
can be regarded as the analytical evolution of a discretised model
of the physics equations, where the computer only provides appreciated
efficiency in the evaluations, but does not add anything new to an
analytical investigation. 

This leads to our point of view that the computer model is not an
independent pillar of science, but an increasingly important contemporary
method in the development of theory. The computer model is most helpful
if many different physics aspects are intertwined in a manner that
prevents the treatment of the problem by dividing it into subproblems.
Classical examples are found in research areas whose subject is inaccessible
to human manipulation and preparation in the laboratory, like meteorology,
planetary interiors, stars, and astrophysical dynamics in general.

Because physics on many different length and time scales is involved,
it is of primary importance to first perform order of magnitude estimates
to identify irrelevant processes and to find equilibrium conditions
that constrain degrees of freedom. A classical theoretical model is
then built by the description of a sequence of dominant processes,
supported by order of magnitude estimates. The next step in the traditional
approach is to approximate the complicated microscopic input physics
by simple analytical fitting formulas that depend on few parameters.
If the fitting formulas are simple enough, the problem might become
analytically solvable so that the dependence of the global features
on the chosen parameters can be revealed. In stellar structure, for
example, order of magnitude arguments show that the matter is in thermodynamical
equilibrium. Hence, an equation of state can be defined that expresses
the gas pressure, $p$, as a function of the local density, $\rho$,
temperature and composition. In some regimes, the microscopic physics
information in the equation of state is well approximated by an equation
of state of the form $p=\kappa\rho^{1+1/n}$. Here, $\kappa$ is a
constant and $n$ a parameter called the polytropic index. If this
approximation holds, the equations of stellar structure can be integrated
analytically for selected integer values of $n$ \cite{1967aits.book.....C,Lane1869}.
Hence, traditional theoretical astrophysics relies
on two prongs: (1) the interpretation of observations in terms of
scenarios that are supported by order of magnitude estimates, and
(2) the explanation of complicated processes by simple approximate
laws that allow a precise analytical investigation of the most important
aspects of the model.

It is obvious how scientific computing can extend the lever of approach
(2). As soon as the complicated processes are reduced to simple approximate
laws, the initial state and the equations of evolution of the model
are mathematically well-defined. It is therefore possible to replace
the former analytical investigation by calculations on the computer.
For example, the Lane-Emden equation can be integrated numerically
for many more values of the polytropic index. There is only one mathematically
correct answer to each problem and the numerical calculation can be
refined until the algorithm obtains the desired precision of the result.

It is much more intricate to let computer models contribute to approach
(1), i.e. the interpretation of observations in terms of scenarios
that are supported by order of magnitude estimates. A well-known example
is the propagation of a shock front. Only a minority of computer codes
attempt to resolve the microscopic width of the shock to correctly
treat its complicated input physics. Most codes just ensure the accurate
conservation of mass, momentum and energy across the shock front to
obtain the shock propagation speed and the thermodynamic conditions
on both sides of the shock. Here, it is the numerical algorithm that
dynamically performs the simplification of the model and not the description
of the input physics like in approach (2). Hence, one may implement
a rich set of input physics and design the finite differencing such
that fundamental laws of physics are fulfilled under all possible
conditions. Then, the complexity of the model is bound by the scale
on which unresolvable small-scale structures are dissipated in space
and time. Because this scale depends on the numerical algorithm and
the mesh topology rather than the investigated physics, different
solutions of the same physics problem may not converge to the same
microscopic solution. However, the goal of approach (1) is not to
evaluate the detailed microscopic state of $\sim10^{9}$ individual
fluid cells (for a human being it would take several years of uninterrupted
reading to absorb this amount of information), but to elaborate the
generic macroscopic properties that are determined by the fundamental
physics laws in combination with the microscopic input physics. The
actual microscopic state of the fluid cells acts as an exemplary sampling
of a specific macroscopic state. Convergence of the evolution of the
macroscopic state must be demonstrated by careful comparison and theoretical
interpretation of concurrent implementations of the same physics problem
using different algorithms and meshes. Hence, we believe that it is
important for the reliability of astrophysical models to be investigated
with several different radiation-magneto-hydrodynamics codes that
are simple and efficient enough to be broadly used.

There are several well-documented and publicly available software
packages. \noun{gadget} is a cosmological N-body code based on the
Smoothed Particle Hydrodynamics method and parallelised using the
Message Passing Interface (MPI). It is a Lagrangean approach with
a hierarchical tree for the non-local evaluation of Newtonian gravity
\cite{2001NewA....6...79S}.
\noun{vh-1} is a multidimensional
hydrodynamics code based on the Lagrangian remap version of the Piecewise
Parabolic Method (PPM) \cite{1993ApJS...88..589B}
that has been further developed and parallelised
using MPI.
\noun{zeus-2D} \cite{1992ApJS...80..819S,1993ApJ...413..198S,1993ApJ...413..210S}
and \noun{zeus-3D} are widely used grid-based hydrodynamics
codes for which a MPI-parallel \noun{zeus-mp} version exists as well.
It offers the choice of different advection schemes on a fixed or
moving orthogonal Eulerian mesh in a covariant description.
While these more traditional approaches distribute in the form of a software
package that includes options to switch on or off, recent open source
projects try to provide the codes in the form of a generic framework
that can host a variety of different modules implementing different
techniques. In this category we could mention the \noun{flash} code
\cite{2002ApJS..143..201C,2000ApJS..131..273F} with its main focus on the
coupling of adaptive mesh stellar hydrodynamics to nuclear burning and
the \noun{whisky} code \cite{2007CQGra..24..235G} as a recent general
relativistic hydrodynamics code based on the \noun{cactus} environment.
Further we mention \noun{athena} \cite{0067-0049-178-1-137} and 
\noun{enzo} \cite{2004astro.ph..3044O} which are both well documented and
publicly available codes.

In this paper we present our code \noun{fish} (Fast and Simple Ideal
magneto-Hydrodynamics), which follows a somewhat different strategy.
\noun{fish} is based on the publicly available serial version of a
cosmological hydrodynamics code \cite{2003ApJS..149..447P}. Its main
virtue is the simplicity and the straightforward approach on a Cartesian
mesh. It can equally well be used for educational purpose in hydrodynamics
classes than for three-dimensional high-resolution astrophysical simulations
\cite{2003ApJ...596L.207P}.
However, in the first code versions, simplicity and parallel efficiency
were not available at the same time. Here we describe a new and elegant
implementation of the parallelisation with a hybrid MPI/OpenMP approach
that has been redesigned from scratch and is well separated from the
subroutines describing the physical conservation equations. Hence,
the same code version is now simple \emph{and} efficient on large
high-performance computing clusters. In comparison to the above-mentioned
software packages it is simple to modify and/or extend and has negligible
overhead. With respect to earlier versions, the discretisation of
the advection terms and the implementation of gravity has further
been developed to make the \noun{fish} code more accurate in the treatment
of fast cold flows and more stable for the evolution of gravitationally
bound hydrostatic objects. Special attention was given to the robustness
of the approach in regions with poor resolution, which are difficult
to avoid in global astrophysical models. \noun{fish} has successfully
been used in one of the first predictions of the gravitational wave
signal from 3D supernova models with microscopic input physics
\cite{2008A&A...490..231S} and provides the foundation for the implementation of
spectral neutrino transport in our new code \noun{elephant} (ELegant
and Efficient Parallel Hydrodynamics with Approximate Neutrino Transport)
\cite{2009ApJ...698.1174L}.

In Section 2 we describe the physics equations and document the numerical
methods used in \noun{fish}. In Section 3 we discuss the optimisation
of the memory access, the parallelisation of \noun{fish} with MPI
and OpenMP and present the scaling behaviour to $\sim10000$ processes.
In Section 4 we finally investigate the performance of \noun{fish}
in a suit of six mostly magneto-hydrodynamic test problems.

\section{Numerical methods}
\label{sec:numerical_methods}
Solving the MHD equations numerically involves overcoming at least two
challenges.
Firstly the difficulty of solving the Riemann problem, used as
building blocks for Godunov type shock capturing numerical schemes, due
to the non-strict hyperbolicity of the MHD equations, see e.g.
{\cite{Barmin1996,LeVeque2005,torrilhon2003}.
The second problem is the divergenceless constraint on the magnetic
field.
A non-vanishing divergence of the magnetic field can produce an acceleration
of the magnetised fluid parallel to the field lines
\cite{1980JCoPh..35..426B}.

The algorithm of Pen et al. \cite{2003ApJS..149..447P} solves the
first problem by using a Riemann solver free relaxation method
\cite{1995CPAM..48..235}.
The second issue is addressed by using a constrained transport method
 \cite{1988ApJ...332..659E} on a staggered grid.

The algorithm of Pen et al. makes extensive use of operator
splitting.
In particular, the evolution of the conducting fluid and the
magnetic field are split by first holding the magnetic field constant
and evolving the fluid and then performing the reverse action.
The source terms arising due to gravity are also accounted for in the
operator splitting method.
Further, the full three dimensional problem is dimensionally split into
one-dimensional subproblems.
An adequately ordered application of the solution operators permits
second order accuracy in time \cite{Strang1968}.

\subsection{The equations of ideal magnetohydrodynamics}
\label{subsec:the_equations_of_ideal_magnetohydrodynamics}
The equations of ideal magnetohydrodynamics (MHD) describe the movement
of a compressible conducting fluid subject to magnetic fields. In ideal MHD
all dissipative processes are neglected, meaning that the fluid possesses
no viscosity and its conductivity is assumed to be infinite. The ideal MHD
equations then read \cite{Landau1991a}
\begin{eqnarray}
\label{eq0010}
  \frac{\partial \rho}{\partial t}
  + \nabla \cdot \left( \rho \boldsymbol{v} \right) & = & 0 \\
\label{eq0020}
  \frac{\partial \rho \boldsymbol{v}}{\partial t}
  + \nabla \cdot \left( 
      \boldsymbol{v} \rho \boldsymbol{v} - \boldsymbol{b} \boldsymbol{b}
    \right)
  + \nabla P_{tot} & = & - \rho \nabla \phi \\
\label{eq0030}
  \frac{\partial E}{\partial t}
  + \nabla \cdot \left[
    \left( E + P_{tot} \right) \boldsymbol{v}
    - \boldsymbol{b} \left( \boldsymbol{v} \cdot \boldsymbol{b} \right)
   \right]
  & = & - \rho \boldsymbol{v} \cdot \nabla \phi \\
\label{eq0040}
  \frac{\partial \boldsymbol{b}}{\partial t}
  - \nabla \times \left( \boldsymbol{v} \times \boldsymbol{b}  \right)
  & = & 0
  ,
\end{eqnarray}
expressing the conservation of mass, momentum, energy and magnetic flux,
respectively.
Here $\rho$ is the mass density, $\boldsymbol{v}$ the velocity and
$E=\rho e+\frac{\rho}{2}v^2+\frac{b^2}{2}$ the total energy being,
the sum of internal, kinetic and magnetic energy.
The magnetic field is given by
$\boldsymbol{B}=/\sqrt{4\pi}\boldsymbol{b}$ and
$P_{tot}=p+\frac{b^2}{2}$ is the total pressure, being the sum of the
gas pressure and the magnetic pressure.
For the equation of state we assume an ideal gas law
\begin{equation}
\label{eq0050}
  p = \rho e ( \gamma - 1)
\end{equation}
where $\gamma$ is the ratio of specific heats.
The right hand side of the momentum and energy conservation equations
detail the effect of gravitational forces onto the conserved variables.
The gravitational potential $\phi$ is determined by the Poisson
equation
\begin{equation}
\label{eq0060}
  \nabla^2 \phi = 4 \pi \rho.
\end{equation}
The MHD equations (\ref{eq0010}-\ref{eq0040}) conserve the divergence
of the magnetic field so that an initial condition
\begin{equation}
\label{eq0061}
  \nabla \cdot \boldsymbol{b} = 0
\end{equation}
remains true, consistent with the physical observation that magnetic
monopoles have never been observed.

Before we start describing the individual solution operators, we first
introduce our notation.
We discretise time into discrete steps $\Delta t^n$ and space into
finite volumes or cells $V_{i,j,k}$ where $n$ labels the different time
levels and the triple $(i,j,k)$ denotes a particular cell.
The vector
$\boldsymbol{u}=\left(\rho,\rho v_x,\rho v_y,\rho v_z,E\right)^T$
denotes the conserved fluid variables.
The solution vector $\boldsymbol{u}^n_{i,j,k}$ contains the spatially
averaged values of the conserved variables at time $t$ in
cell $V_{i,j,k}$
\begin{equation}
\label{eq0070}
  \boldsymbol{u}_{i,j,k}
  = \frac{1}{V_{i,j,k}} \int_{V_{i,j,k}}
    \boldsymbol{u} \left( \boldsymbol{x},t \right) \ud x \ud y \ud z,
\end{equation}
where the cell volume $V_{i,j,k}=\Delta x\Delta y\Delta z$ is given by
the assumed constant cell dimensions
$\Delta x=x_{i'}-x_{i'-1}$, $\Delta y=y_{j'}-y_{j'-1}$,
$\Delta z=z_{k'}-z_{k'-1}$.
Half-integer indices are indicated by a prime $i'=i+1/2$, $j'=j+1/2$,
$k'=k+1/2$ and denote the intercell boundary.
Further we define the cell face averaged magnetic field components at
time $t$ by
\begin{equation}
\label{eq0080}
  \left( b_x \right)_{i',j,k}
  = \frac{1}{S_{i',j,k}}
    \int_{S_{i',j,k}} b_x \left( \boldsymbol{x},t \right) \ud y \ud z
\end{equation}
\begin{equation}
\label{eq0080a}
  \left( b_y \right)_{i,j',k}
  = \frac{1}{S_{i,j',k}}
    \int_{S_{i,j',k}} b_y \left( \boldsymbol{x},t \right) \ud x \ud z
\end{equation}
\begin{equation}
\label{eq0080b}
  \left( b_z \right)_{i,j,k'}
  = \frac{1}{S_{i,j,k'}}
    \int_{S_{i,j,k'}} b_z \left( \boldsymbol{x},t \right) \ud x \ud y
\end{equation}
where $S_{i',j,k}=\Delta y\Delta z$ denotes the cell face of cell
$V_{i,j,k}$ located at $x_{i'}$ and spanned by the zone increments
$\Delta y$ and $\Delta z$.
$\left( b_y \right)^n_{i,j',k}$ and $\left( b_z \right)^n_{i,j,k'}$
are defined analogously.

In an operator-split scheme the solution algorithm to the ideal MHD
equations can be summarized as
\begin{equation}
\label{eq0081}
  \boldsymbol{u}^{n+2} =
    L_{\mathrm{forward}} L_{\mathrm{backward}} \boldsymbol{u}^{n}
  ,
\end{equation}
where
\begin{equation}
\label{eq0082}
  \begin{aligned}
  L_{\mathrm{forward}}  & =
    L_x\left(\Delta t\right) B^{yz}_x\left(\Delta t\right)
    L_y\left(\Delta t\right) B^{xz}_y\left(\Delta t\right)
    L_z\left(\Delta t\right) B^{xy}_z\left(\Delta t\right) \\
  L_{\mathrm{backward}} & =
    L_z\left(\Delta t\right) B^{xy}_z\left(\Delta t\right)
    L_y\left(\Delta t\right) B^{xz}_y\left(\Delta t\right)
    L_x\left(\Delta t\right) B^{yz}_x\left(\Delta t\right)
  \end{aligned}
\end{equation}
are the forward and backward operator for one time step.
The operators $L_{x,y,z}$ evolve the fluid and
account for the source terms, while the $B$ operators evolve the
magnetic field.
If the individual operators are second order accurate, then the
application of the forward followed by the backward operator is
second order accurate in time \cite{Strang1968}.
In the following subsections we shall detail the individual operators.

The numerical solution algorithm to the MHD equations is explicit.
Hence we are restricted by the Courant, Friedrich and Lewy
\cite{Courant1928} (CFL) condition.
Therefore we impose the following time step
\begin{equation}
\label{eq0083}
  \Delta t^n = k \cdot
      \min_{i,j,k} \left( \frac{\Delta x}{C_{i,j,k}^{n,x}}
                         ,\frac{\Delta y}{C_{i,j,k}^{n,y}}
                          ,\frac{\Delta z}{C_{i,j,k}^{n,z}} \right)
\end{equation}
where
\begin{equation}
\label{eq0084}
  C_{i,j,k}^{n,d} = 
    \max \left(
      v_{d,i,j,k}^n + c_{F i,j,k}^n
    \right)
\end{equation}
is the maximum speed at which information can travel in the whole
computational domain in direction $d=x,y,z$ being the sum of the
velocity component in d and the speed of the fast magnetosonic waves
$c_F$.
We typically set the CFL number $k$ to $0.75$.

\subsection{Solving the fluid MHD equations}
\label{subsec:solvingthefluidmhdequations}
In this subsection we describe the evolution of the fluid variables
$\boldsymbol{u}$ in the $x$-direction.
During this process the magnetic field is held constant and interpolated
to cell centers. The gravitational source
terms are assumed to vanish for the moment. Then mass, momentum and
energy conservation in $x$ direction can be written as
\begin{equation}
\label{eq0090}
    \frac{\partial \boldsymbol{u}}{\partial t}
  + \frac{\partial \boldsymbol{F}}{\partial x} = 0
\end{equation}
where
\begin{equation}
\label{eq0100}
  \begin{aligned}
  \boldsymbol{F} = \left[ \begin{array}{c}
    \rho v_x \\
    \rho v_x^2 + P_{tot} - b_x^2 \\
    \rho v_x v_y - b_x b_y \\
    \rho v_x v_z - b_x b_z \\
    (E + P_{tot}) v_x - b_x \boldsymbol{b} \cdot \boldsymbol{v}
  \end{array}\right]
  \end{aligned}
\end{equation}
is the flux vector.

Integrating eq. \eqref{eq0090} over a cell $V_{i,j,k}$ gives
\begin{equation}
\label{eq0101}
    \frac{\partial \boldsymbol{u}_{i,j,k}}{\partial t}
  + \frac{1}{\Delta x} \left(
    \boldsymbol{F}_{i',j,k} - \boldsymbol{F}_{i'-1,j,k}
  \right) = 0
\end{equation}
where the definition of the cell averaged values \eqref{eq0070} has
been substituted and Gauss' theorem has been used.
The numerical flux $\boldsymbol{F}_{i',j,k}$ represents an average
flux of the conserved quantities through the surface $S_{i',j,k}$
\begin{equation}
\label{eq0102}
  \boldsymbol{F}_{i',j,k}
    = \frac{1}{S_{i',j,k}}
      \int_{S_{i',j,k}} 
      \boldsymbol{F} \left( \boldsymbol{x},t \right) \ud y \ud z
\end{equation}
at given time $t$. Eq. \eqref{eq0101} is a semi-discrete conservative
scheme for the conservation law \eqref{eq0090}.
In the following we focus on obtaining the numerical fluxes in a stable
and accurate manner. Time integration of the ordinary differential
equation \eqref{eq0101} will be addressed later in this subsection.

Many schemes for the stable and accurate computation of the numerical
fluxes have been devised in the literature.
Godunov type methods achieve this by solving either exact or
approximate Riemann problems at cell interfaces
\cite{Godunov1959,Laney1998,Toro1997}.
Through solving the Riemann problem, these methods ensure an upwind
discretisation of the conservation law and hence achieve causal
consistency.
Due to the already mentioned difficulty of solving the
Riemann problem in the ideal MHD case, the algorithm of Pen at el.
\cite{2003ApJS..149..447P} uses the relaxation scheme of Jin and Xin
\cite{1995CPAM..48..235}.
For detailed information on these type of methods we refer to
\cite{1995CPAM..48..235,LeVeque2001} and the references therein.

The idea of the relaxation scheme is to replace a system like
\eqref{eq0100} by a larger system
\begin{equation}
\label{eq0110}
  \begin{aligned}
  \frac{\partial \boldsymbol{u}}{\partial t}
    + \frac{\partial \boldsymbol{w}}{\partial x} & = 0 \\
  \frac{\partial \boldsymbol{w}}{\partial t}
  + D^2 \frac{\partial \boldsymbol{u}}{\partial x} 
    & = \frac{1}{\epsilon}
    \left( \boldsymbol{F}(\boldsymbol{u}) - \boldsymbol{w} \right)
  ,
  \end{aligned}
\end{equation}
called the relaxation system.
Here, the relaxation rate $\epsilon$ is a small positive
parameter and $D^2$ is a positive definite matrix.
For small relaxation rates, system \eqref{eq0110} rapidly relaxes to
the local equilibrium defined by
$\boldsymbol{w}=\boldsymbol{F}(\boldsymbol{u})$.
A necessary condition for solutions of the relaxation
system \eqref{eq0110} to converge in the small $\epsilon$ limit to
solutions of the original system \eqref{eq0090} is that the
characteristic speeds of the hyperbolic part of \eqref{eq0110} are at
least as large or larger than the characteristic speeds in
system \eqref{eq0090}.
This is the so-called subcharacteristic condition.

As in Ref. \cite{1995CPAM..48..235} we choose $D=d\cdot I$ to be a diagonal
matrix.
In order to fulfill the subcharacteristic condition the
diagonal element $d$ or the so-called freezing speed is chosen to be
\begin{equation}
\label{0111}
  d = | v_x | + c_F
  ,
\end{equation}
where $c_F$ is the speed of the fast magnetosonic waves, i.e. the
fastest wave propagation speed supported by the equations of
ideal MHD.

The key point in the relaxation system is that in the local equilibrium
limit it has a very simple characteristic structure
\begin{equation}
\label{eq0120}
  \begin{aligned}
    \frac{\partial}{\partial t} \left( \boldsymbol{w} + D \boldsymbol{u} \right)
      + D \frac{\partial}{\partial x} \left( \boldsymbol{w} + D \boldsymbol{u} \right) = 0 \\
    \frac{\partial}{\partial t} \left( \boldsymbol{w} - D \boldsymbol{u} \right)
      - D \frac{\partial}{\partial x} \left( \boldsymbol{w} - D \boldsymbol{u} \right) = 0
    ,
  \end{aligned}
\end{equation}
where $\boldsymbol{w} \pm D \boldsymbol{u}$ are then the characteristic
variables. They travel with the ``frozen'' characteristic speeds $\pm D$
respectively.

System \eqref{eq0120} can be easily recast into an equation for
$\boldsymbol{u}$ and $\boldsymbol{w}$.
However, we are practically only interested in that for
$\boldsymbol{u}$
\begin{equation}
\label{eq0130}
  \frac{\partial \boldsymbol{u}}{\partial t}
  + \frac{\partial \boldsymbol{F^+}}{\partial x}
  + \frac{\partial \boldsymbol{F^-}}{\partial x}
  = 0
\end{equation}
where $\boldsymbol{F}^+=(\boldsymbol{w} + D \boldsymbol{u})/2$ denotes
the right travelling waves and
$\boldsymbol{F}^-=(\boldsymbol{w} - D \boldsymbol{u})/2$ the left
travelling waves.
In the follwing we shall drop the indices of the ignored directions.
Since this defines an upwind direction for each wave component, a first
order upwind scheme results from choosing
$\boldsymbol{F}^+_{i'}=\boldsymbol{F}^+_{i}$ and
$\boldsymbol{F}^-_{i'}=\boldsymbol{F}^-_{i+1}$.
In this case, the total flux at the cell interfaces is readily
evaluated to become
\begin{equation}
\label{eq0140}
  \boldsymbol{F}_{i'} = \boldsymbol{F}^+_{i'} + \boldsymbol{F}^-_{i'}
  = \frac{1}{2} \left( \boldsymbol{F}_i + \boldsymbol{F}_{i+1} \right)
  - \frac{1}{2} D \left( \boldsymbol{u}_{i+1} - \boldsymbol{u}_{i} \right)
\end{equation}
where $\boldsymbol{F}_i=\boldsymbol{w}_i=\boldsymbol{F}(\boldsymbol{u}_{i})$.
For $D$ we use the freezing speed
\begin{equation}
\label{eq0150}
  d = \max \left( d_{i} , d_{i+1} \right)
\end{equation}
in order to satisfy the subcharacteristic condition.
We note that this choice for the freezing speed makes the numerical flux
equivalent to the Rusanov flux and the local Lax-Friedrichs flux.
As pointed out in \cite{LeVeque2001}, a wide variety of numerical flux
assignments can be derived from the relaxation system by simply letting
the matrix $D$ having a more complicated form than diagonal.

So far, the numerical flux \eqref{eq0140} is only first order accurate.
First order methods permit the automatic capturing of flow discontinuities
but are inaccurate in smooth flow regions due to the large
amount of numerical dissipation inherent to them.
As a matter of fact, the large numerical dissipation present in first
order methods is not a deficit of theses methods but it is the reason
why they are stable at flow discontinuities in the first place.
However, in many applications both smooth and discontinuous flow
features are present and therefore the use of higher order methods is
desirable.
We opt for a second order accurate total variation diminishing (TVD)
scheme due to the low computational cost and the robustness of these
type of schemes.

Let us first consider the right traveling waves $\boldsymbol{F}^+$.
Given the $i$th cell, a first order accurate flux at the cell boundary
$x_{i'}$ is then given by
$\boldsymbol{F}^+_{i'}=\boldsymbol{F}^+_i=\boldsymbol{F}^+(\boldsymbol{u}_i)$.
This corresponds to a piece-wise constant approximation of the flux
function $\boldsymbol{F}^+(x,t)$ over the staggered cell
$[x_i,x_{i+1}]$.
For second order accuracy we seek a piece-wise linear approximation
\begin{equation}
\label{eq0160}
  \boldsymbol{F}^+(x,t) \approx \boldsymbol{F}^+_i
    + \left. \frac{\partial \boldsymbol{F}^{+}}{\partial x} \right|_i (x - x_i)
\end{equation}
where the derivative may be approximated from first order flux
differences.
Two choices exist: either left or right differences
\begin{equation}
\label{eq0170}
  \left. \frac{\partial \boldsymbol{F}^{+}}{\partial x} \right|_i
  = \left\{
    \begin{array}{l l l}
      \Delta \boldsymbol{F}^{+,L}_{i}
      = \left( \boldsymbol{F}^{+}_{i} -  \boldsymbol{F}^{+}_{i-1} \right)
      / \Delta x \\ \\
      \Delta \boldsymbol{F}^{+,R}_{i}
      = \left( \boldsymbol{F}^{+}_{i+1} -  \boldsymbol{F}^{+}_{i} \right)
      / \Delta x \\
    \end{array}
  \right.
  .
\end{equation}
To choose between the left $\Delta \boldsymbol{F}^{+,L}_{i}$ and
right $\Delta \boldsymbol{F}^{+,R}_{i}$ differences a flux limiter
$\phi$
\begin{equation}
\label{eq0180}
  \Delta \boldsymbol{F}^{+}_{i}
  = \phi \left( \Delta \boldsymbol{F}^{+,L}_{i},\Delta \boldsymbol{F}^{+,R}_{i} \right)
\end{equation}
is used.

This limiter enforces a nonlinear stability constraint commonly known
as TVD to ensure the stability of the scheme in the vicinity of
discontinuities. The limiter reduces spurious oscillations associated
with higher accuracy than first order to get a high resolution method.
See for example \cite{Laney1998,LeVeque1992,LeVeque2005,Toro1997}
and references therein.

We have implemented the minmod limiter
\begin{equation}
\label{eq0190}
  \phi(a,b) = \mathrm{minmod}(a,b)
  = \frac{1}{2} \left( \mathrm{sign}(a) + \mathrm{sign}(b) \right) \min ( |a|,|b| )
\end{equation}
which chooses the smallest absolute difference if both have the same
sign, and the Van Leer limiter
\begin{equation}
\label{eq0200}
  \phi(a,b)
  = \frac{1}{2} \left( \mathrm{sign}(a) + \mathrm{sign}(b) \right)
    \frac{2ab}{a+b}
  .
\end{equation}
Other choices are possible for the scheme to be TVD
\cite{1995CPAM..48..235}.
Note that when the left and right flux differences have different
signs, i.e. at extrema and hence also at shocks, no correction
is added in \eqref{eq0160} and the scheme switches to first order accuracy.
For core-collapse simulations we use the Van Leer limiter in the
subsonic flow regions and the minmod limiter in supersonic
regions.

In a similar way we may construct a piece-wise linear approximation for
the left going fluxes $\boldsymbol{F}^{-}$ in the staggered cell
$[x_i,x_{i+1}]$ starting at $x_{i+1}$
\begin{equation}
\label{eq0201}
  \boldsymbol{F}^-(x,t) \approx \boldsymbol{F}^-_{i+1}
    + \left. \frac{\partial \boldsymbol{F}^{-}}{\partial x} \right|_{i+1} (x - x_{i+1})
\end{equation}
with either the left or right differences
\begin{equation}
\label{eq0210}
  \left. \frac{\partial \boldsymbol{F}^{-}}{\partial x} \right|_{i+1}
  = \left\{
    \begin{array}{l l l}
      \Delta \boldsymbol{F}^{-,L}_{i+1}
      = \left( \boldsymbol{F}^{-}_{i+1} -  \boldsymbol{F}^{-}_{i} \right)
      / \Delta x \\ \\
      \Delta \boldsymbol{F}^{-,R}_{i+1}
      = \left( \boldsymbol{F}^{-}_{i+2} -  \boldsymbol{F}^{-}_{i+1} \right)
      / \Delta x \\
    \end{array}
  \right.
  .
\end{equation}
Again the flux limiter $\phi$ is used to discriminate between the left
or right differences
\begin{equation}
\label{eq0220}
  \Delta \boldsymbol{F}^{-}_{i+1}
  = \phi \left( \Delta \boldsymbol{F}^{-,L}_{i+1},\Delta \boldsymbol{F}^{-,R}_{i+1} \right).
\end{equation}
The total second order accurate numerical flux is then simply
\begin{equation}
\label{eq0230}
  \boldsymbol{F}_{i'} = \boldsymbol{F}^{+}_{i} + \boldsymbol{F}^{-}_{i+1}
  + \frac{\Delta x}{2} \left( \Delta \boldsymbol{F}^{+}_{i}
     - \Delta \boldsymbol{F}^{-}_{i+1} \right).
\end{equation}

For the time integration of eq. \eqref{eq0101} we use a two step
predictor-corrector method.
As predictor we compute a half time step with the first order fluxes
\eqref{eq0140}.
We regard the freezing speed in the predictor step as a parameter
varying between $d=0$ and $d=\max\left(d_{i},d_{i+1}\right)$ to
regulate the numerical dissipation.
Hence we vary the predictor between a first order scheme and a second
order centered difference scheme depending on the application.

In the corrector step we then use the calculated values from the
predictor step $\boldsymbol{u}^{n'}$ to compute the second order TVD
fluxes eq. \eqref{eq0230}:
\begin{equation}
\label{eq0240}
  \boldsymbol{u}_{i}^{n+1} = \boldsymbol{u}_{i}^{n}
  - \frac{\Delta t}{\Delta x} \left(
    \boldsymbol{F}_{i'}^{n'} - \boldsymbol{F}_{i'-1}^{n'}
  \right)
  .
\end{equation}
Hence we obtain a second order update in time and space of the fluid
variables.
This ends the description of the $L_x$ solution operator.
The other spatial directions are treated in the same way.

\subsection{Incorporation of gravitation}
\label{subsec:incorparation_of_gravitation}
Gravitational forces play an important role in most astrophysical
processes.
A code devoted to the simulation of these processes should therefore
include gravity and we describe our implementation in the following
subsection.
To integrate the source terms, we split the source term
dimensionally as follows:
\begin{equation}
\label{eq0251}
    \frac{\partial \boldsymbol{u}}{\partial t}
  + \frac{\partial \boldsymbol{F}}{\partial x} = \boldsymbol{S}_x
\end{equation}
where
$\boldsymbol{S}_x=(0,-\rho \partial \phi / \partial x,0,0,-\rho v_x \partial \phi / \partial x)$,
and in an analogous manner for the $y$ and $z$ direction.
In the following we shall regard the gravitational potential as given
and constant in time.
For the integration of eq. \eqref{eq0251} one then has two
possibilities, either an operator split or an unsplit method.

In the fully operator split version, the evolution of the conserved
variables is divided into a homogeneous system ($\boldsymbol{S}_x=0$)
and the ordinary differential equation
\begin{equation}
\label{eq02511}
    \frac{\ud \boldsymbol{u}}{\ud t} = \boldsymbol{S}_x
\end{equation}
Solving the homogeneous part has been discussed in the previous
section.
In operator notation, we then solve equation \eqref{eq0251} as
\begin{equation}
\label{eq025111}
  \boldsymbol{u}^{n+1} =
  G_x \left(\frac{\Delta t}{2}\right)
  L_x \left(\Delta t\right)
  G_x \left(\frac{\Delta t}{2}\right)
\end{equation}
which is second order accurate in time.

Further, for reasons to be explained in section
\ref{sec:parallelisation_and_performance}, our code has as fundamental
variables density, momentum and internal energy.
Therefore we only update the momentum field, and the operator
$G_x$ is then explicitly
\begin{equation}
\label{eq02512}
  G_x\left(\Delta t\right) :
  \;
  (\rho\boldsymbol{v})_i^{n+1} = (\rho\boldsymbol{v})_i^{n}
  - \Delta t \cdot
  \rho_i^{n} \left( \frac{\partial \phi}{\partial x} \right)_i^n
  ,
\end{equation}
where we use centered differences for the gravitational potential
\begin{equation}
\label{eq02513}
  \left( \frac{\partial \phi}{\partial x} \right)_i^n
  = \frac{\phi_{i+1}^n - \phi_{i-1}^n}{2 \Delta x}
  .
\end{equation}
Note that the density field is left constant according to \eqref{eq02511}.
The total energy is computed by summing up internal energy, magnetic
energy and kinetic energy and given as input to the $L_x$
operator.
Therefore the total energy change due to the source term is implied from
the updated momentum field.

As a second possibility, we implemented an unsplit version.
There we directly account for gravity in the fluid
predictor/corrector steps.
The predictor step then is given by
\begin{equation}
\label{eq0253}
  \boldsymbol{u}_{i}^{n'} = \boldsymbol{u}_{i}^{n}
  - \frac{\Delta t}{2 \Delta x} \left(
    \boldsymbol{F}_{i'}^{n} - \boldsymbol{F}_{i'-1}^{n}
  \right) + \frac{\Delta t}{2} \boldsymbol{S}_i^{n}
\end{equation}
where the $\boldsymbol{F}_{i'}^{n}$ is the predictor
flux as described in previous section and
\begin{equation}
\label{eq0254}
  \begin{aligned}
  \boldsymbol{S}_i^{n} = \left[ \begin{array}{c}
    0 \\
    \rho_i^n \\
    0 \\
    0 \\
    (\rho v_x)_i^n \\
  \end{array}\right]
  \frac{\phi_{i+1} - \phi_{i-1}}{2 \Delta x}
  \end{aligned}.
\end{equation}
The corrector step is then
\begin{equation}
\label{eq0255}
  \boldsymbol{u}_{i}^{n+1} = \boldsymbol{u}_{i}^{n}
  - \frac{\Delta t}{\Delta x} \left(
    \boldsymbol{F}_{i'}^{n'} - \boldsymbol{F}_{i'-1}^{n'}
  \right) + \Delta t \boldsymbol{S}_i^{n'}
  ,
\end{equation}
where the fluid fluxes are given in previous section and
$\boldsymbol{S}_i^{n'}$ is analogous to \eqref{eq0254}.
However, the density and the momentum are then given by the
predictor step $n'$.

Both implementations of the source terms are second order
accurate in space and time.

Astrophysical simulations often include gravitationally bound objects
that can be in hydrostatic equilibrium (HSE).
With the described discretisation of the fluxes, the gradients due
to the hydrostatic stratification are interpreted by the
fluid scheme as a propagating wave and the scheme invokes numerical
dissipation.
In HSE, however, the gradients are not the result of a propagating wave
but rather a consequence of gravity which is stationary.
In order to subtract the gravitationally induced gradients from the
algorithm that tailors the dissipation in the TVD scheme, we try to
find an analytical estimate of the gradients present in HSE.

In the following we neglect the influence of magnetic fields and
are considering only gas pressure supported equilibrium.
In HSE, the pressure variation is then given by
\begin{equation}
\label{eq0260}
  \frac{1}{\rho} \frac{\partial p}{\partial x}
  = - \frac{\partial \phi}{\partial x}
  .
\end{equation}
Furthermore we neglect entropy and composition gradients.
The later assumption is needed for more complex equations of state
$p=p(\rho,s,\boldsymbol{Y})$ where $\boldsymbol{Y}$ denotes
a vector of chemical compositions.
Under these assumptions, we can express the pressure gradient as
\begin{equation}
\label{eq0270}
  \frac{\partial p}{\partial x}
  = \left( \frac{\partial p}{\partial \rho} \right)_{s,\boldsymbol{Y}} \frac{\partial \rho}{\partial x}
  = c_S^2 \frac{\partial \rho}{\partial x}
  ,
\end{equation}
where $c_S$ is the speed of sound.
Then the density gradient in HSE can be expressed with \eqref{eq0260}
as
\begin{equation}
\label{eq0280}
  \frac{\partial \rho}{\partial x}
  = - \frac{\rho}{c_S^2} \frac{\partial \phi}{\partial x}
  .
\end{equation}
If we now relax the HSE condition by allowing the velocity to be non
zero but roughly constant across a few cells of investigation, we can
rewrite the momentum gradient by using the continuity
equation
\begin{equation}
\label{eq0290}
  \frac{\partial \rho v_x}{\partial x}
  = - \frac{\rho v_x}{c_S^2} \frac{\partial \phi}{\partial x}
  .
\end{equation}
Similar equations can be derived for the $y$ and $z$ components of
the momentum gradient in $x$ direction.
Using the fundamental thermodynamic relation $\ud e=-p\ud V$ an energy
gradient can be derived as well
\begin{equation}
\label{eq0300}
  \frac{\partial E}{\partial x}
  = - \frac{1}{c_S^2} \left( E + p \right) \frac{\partial \phi}{\partial x}
  .
\end{equation}
In summary, we then have the following gradients in HSE
\begin{equation}
\label{eq0310}
  \frac{\partial \boldsymbol{u}}{\partial x}
  = \frac{\boldsymbol{\xi}}{c_{S}^2} \frac{\partial \phi}{\partial x}
  ,
\end{equation}
where
\begin{equation}
\label{eq0320}
  \boldsymbol{\xi} = -
  \left(  \begin{array}{c}
    \rho \\
    \rho v_x \\
    \rho v_y \\
    \rho v_z \\
    E + p
  \end{array} \right)
  .
\end{equation}

This estimate of the gradients in HSE are now subtracted from the
difference $\boldsymbol{u}_{i+1}-\boldsymbol{u}_i$, that generates
the numerical dissipation when it is multiplied by the freezing speed.
First we multiply \eqref{eq0310} by $\ud x$ and discretise the result
as
\begin{equation}
\label{eq0321}
  \boldsymbol{g}_i
  = \frac{1}{c_{S,i}\cdot c_{S,i+1}}
    \frac{\boldsymbol{\xi}_{i+1}+\boldsymbol{\xi}_i}{2} ( \phi_{i+1}-\phi_i )
  .
\end{equation}
This approximation is more in the spirit of finite differences than
finite volume methods.
Nevertheless, since we only seek second order accuracy we did not
explore more sophisticated methods.
Then we define
\begin{eqnarray}
\label{eq0323}
  \boldsymbol{\Delta u}_i & = & \boldsymbol{u}_{i+1}-\boldsymbol{u}_i \\
  \boldsymbol{\delta u}_i & = & \textrm{minmod}
    \left( \boldsymbol{\Delta u}_i - \boldsymbol{g}_{i}, \boldsymbol{\Delta u}_i \right)
  .
\end{eqnarray}
The minmod function prevents $\boldsymbol{\delta u}_i$ of becoming
antidissipative.
The HSE corrected difference is then used to modify the first order
upwind flux \eqref{eq0140} as
\begin{equation}
\label{eq0324}
  \boldsymbol{F}_{i'}
  = \frac{1}{2} \left( \boldsymbol{F}_i + \boldsymbol{F}_{i+1} \right)
  - \frac{1}{2} D_{\mathrm{adv}} \boldsymbol{\Delta u}_i
  - \frac{1}{2} D_{\mathrm{ac}} \boldsymbol{\delta u}_i
  ,
\end{equation}
where
\begin{eqnarray}
\label{eq0325}
  & D_{\mathrm{adv}} & = \max \left( v_{x,i} , v_{x,i+1} \right) I \\
\label{eq0326}
  & D_{\mathrm{ac}} & = \max \left( c_{S,i} , c_{S,i+1} \right) I
  .
\end{eqnarray}
Hence for the advective part of the numerical dissipation we use the
full difference $\boldsymbol{\Delta u}_i$ while for the acoustic part
we use the HSE corrected difference $\boldsymbol{\delta u}_i$.
The second order fluxes are constructed in the same way as described
in section \ref{subsec:solvingthefluidmhdequations} but with the
HSE corrected first order fluxes.

We note that the previous description of HSE is only directly applicable
for purely hydrodynamic simulations.
However, we have found that by replacing the acoustic speed by the
fast magnetosonic speed
\begin{eqnarray}
\label{eq03261}
  & D_{\mathrm{ac}} & = \max \left( c_{F,i} , c_{F,i+1} \right) I
\end{eqnarray}
works in a stable manner for MHD simulations as well.

For completeness we outline how the HSE correction could be applied
more generally to other flux assignment methods in the context of
purely hydrodynamic simulations, e.g. the method of
Roe.
In that method, the intercell flux is given in the form
\begin{equation}
\label{eq0327}
  \boldsymbol{F}_{i'} =
  \frac{1}{2} \left( \boldsymbol{F}_i + \boldsymbol{F}_{i+1} \right)
  - \frac{1}{2} \sum_{p=1}^{5} \alpha_p | \lambda_p | \boldsymbol{K}_p
  ,
\end{equation}
where the $\lambda_p$ are the eigenvalues and the $\boldsymbol{K}_p$
are the right eigenvectors of the Jacobian matrix of the flux function
$\boldsymbol{\ud F}/\boldsymbol{\ud u}$.
The eigenvalues are $\lambda_1=v_x-c_S$, $\lambda_{2,3,4}=v_x$ and
$\lambda_5=v_x+c_S$.
The eigenvectors can be found, for example, in \cite{Toro1997}.
The coefficients $\alpha_p$ are obtained by projecting the differences
in the conserved variables onto the right eigenvectors:
\begin{equation}
\label{eq0328}
  \boldsymbol{\Delta u}_i
  = \sum_{p=1}^{5} \alpha_p \boldsymbol{K}_p
  .
\end{equation}
The HSE correction can now be used to modify the projection of the
acoustic modes 1 and 5.
Instead of projecting the acoustic modes onto $\boldsymbol{\Delta u}_i$
, one can project them onto the HSE corrected differences
$\boldsymbol{\delta u}_i$.
The advective modes 2,3 and 4 remain identical.

We mention that Ref. \cite{2002ApJS..143..539Z} derived a simple
``modified states'' version of the PPM method with the same aim to
improve the numerical solution in HSE.
However, their method is only directly applicable to the PPM
method.

Alternatively, the gravitational source terms can be completely avoided
by using a fully conservative form of the hydrodynamics equations
(see \cite{1998ApJS..115...19P}).

\subsection{Advection of the magnetic field}
\label{subsec:advection_of_the_magnetic_field}
In this subsection we describe the advection of the magnetic field
in $x$-direction $B_x^{yz}$.
The operators for the update in $y$ and $z$ directions $B_y^{xz}$ and
$B_z^{xy}$ are handled analogously.
During this operator split update, all quantities other than the
magnetic field are held constant.
The update is then prescribed by the induction equation
\begin{equation}
\label{eq0330}
  \frac{\partial \boldsymbol{b}}{\partial t}
  - \nabla \times \left( \boldsymbol{v} \times \boldsymbol{b}  \right)
  = 0.
\end{equation}
Straight forward discretisation of \eqref{eq0330} can only guarantee
that $\nabla\cdot\boldsymbol{b}=0$ is of the order of the truncation
error. However, at flow discontinuities, the discrete divergence may
become large.
As a consequence, large errors in the simulation can accumulate
\cite{Brackbill1980}.
Therefore care has to be taken.
A variety of methods have been proposed to surmount this difficulty,
see e.g.
\cite{1988ApJ...332..659E,Londrillo2004,rossmanith:1766,torrilhon:1694,2000JCoPh.161..605T}.
The algorithm of Pen et. al \cite{2003ApJS..149..447P}, and hence our
code, uses the constrained transport method \cite{1988ApJ...332..659E}.

The key idea of the constrained transport method is to write the
induction equation in integral form.
Integrating eq. \eqref{eq0330}, for example, over the surface
$S_{i',j,k}$ of cell $V_{i,j,k}$, substituting definition
\eqref{eq0080} and using Stoke's theorem yields
\begin{equation}
\label{eq0340}
  \frac{\partial}{\partial t} \left( b_x \right)_{i',j,k}
  = \int_{\partial S_{i',j,k}} \boldsymbol{v} \times \boldsymbol{b}
  \cdot \ud \boldsymbol{x}
\end{equation}
where $\partial S_{i',j,k}$ denotes the contour of $S_{i',j,k}$,
i.e. the edges of the cell-face at $i'$.
The integral form then naturally suggests one to choose the normal
projections of the magnetic field at faces
of the cell $V_{i,j,k}$ and the normal projections of the electric field
$\boldsymbol{E}=\boldsymbol{v} \times \boldsymbol{b}$ at the cell edges
as primary variables.
This positioning leads directly to the jump conditions of electric and
magnetic fields \cite{Jackson1998} and therefore mimics Maxwells
equations at the discrete level.
The discrete form of the $\nabla\cdot\boldsymbol{b}=0$ constraint is
then defined as
\begin{equation}
\label{eq0341}
  \left( \nabla\cdot\boldsymbol{b} \right)^n_{i,j,k}
  = \frac{\left( b_x \right)^n_{i',j,k} - \left( b_x \right)^n_{i'-1,j,k}}{\Delta x}
  + \frac{\left( b_y \right)^n_{i,j',k} - \left( b_y \right)^n_{i,j'-1,k}}{\Delta y}
  + \frac{\left( b_z \right)^n_{i,j,k'} - \left( b_z \right)^n_{i,j,k'-1}}{\Delta z}
  .
\end{equation}

A detailed inspection of the characteristic structure of the induction
equation reveals the presence of two transport modes and one constraint
mode.
As pointed out by Pen et al. \cite{2003ApJS..149..447P}, this allows
one to separate the evolution of the induction equation into advection
and constraint steps.
In $x$-direction, for example, this means that the $y$ and $z$
components of the magnetic field need to be updated as
\begin{equation}
\label{eq0350}
  \begin{aligned}
    \frac{\partial}{\partial t} \left( b_y \right)_{i,j',k}
    & = - \frac{1}{\Delta x}
      \left[
      \left( v_x b_y \right)_{i',j',k}-\left( v_x b_y \right)_{i'-1,j',k}
      \right]
    \\
    \frac{\partial}{\partial t} \left( b_z \right)_{i,j,k'}
    & = + \frac{1}{\Delta x}
      \left[
      \left( v_x b_z \right)_{i',j,k'}-\left( v_x b_z \right)_{i'-1,j,k'}
      \right].
  \end{aligned}
\end{equation}
However, the $x$ component of the magnetic field has to be updated as
\begin{eqnarray}
\label{eq0360}
  \begin{aligned}
  \frac{\partial}{\partial t} \left( b_x \right)_{i',j,k}
  = & + \frac{1}{\Delta y}
      \left[
      \left( v_x b_y \right)_{i',j',k}-\left( v_x b_y \right)_{i',j'-1,k}
      \right]
   \\
    & - \frac{1}{\Delta z}
      \left[
      \left( v_x b_z \right)_{i',j,k'}-\left( v_x b_z \right)_{i',j,k'-1}
      \right].
  \end{aligned}
\end{eqnarray}
The fluxes in eq. \eqref{eq0350} need to be upwinded for stability,
since they represent the two advection modes.
To maintain $\nabla\cdot\boldsymbol{b}=0$ within machine precision, the
same fluxes used to update $b_y$ and $b_z$ in eq. \eqref{eq0350} need to
be used in eq. \eqref{eq0360} for the $b_x$ update.
A simple calculation then clearly shows that
$\partial/\partial t (\nabla\cdot\boldsymbol{b})=0$.

To update $\left( b_y \right)_{i,j',k}$ we then proceed as follows.
Note that the velocity $v_x$ needs to be interpolated to the same
location as $\left( b_y \right)_{i,j',k}$, i.e. to cell face
$S_{i,j',k}$:
\begin{equation}
\label{eq0361}
  (v_x)_{i,j',k} = \frac{(v_x)_{i,j,k}+(v_x)_{i,j+1,k}}{2}
  .
\end{equation}
This simple interpolation is, however, not numerically stable due
to the negligence of causal consistency (or equivalently numerical
dissipation).
It turns out, that for stability a simple averaging in $x$ direction
\begin{equation}
\label{eq0362}
  (v_x)_{i,j',k} = \frac{1}{4}
  \left[    \left(v_x\right)_{i+1,j',k}
        + 2 \left(v_x\right)_{i,j',k}
        +   \left(v_x\right)_{i-1,j',k} \right]
\end{equation}
introduces enough numerical dissipation \cite{2003ApJS..149..447P}.
A first order accurate upwinded flux is then given by
\begin{equation}
\label{eq0363}
  \left( v_x b_y \right)_{i',j',k}
  = \left\{
    \begin{array}{l l l}
      \left( v_x b_y \right)_{i,j',k}   & , \; & (v_x)_{i',j',k} > 0 \\
      \\
      \left( v_x b_y \right)_{i+1,j',k} & , \; & (v_x)_{i',j',k} \leq 0 \\
    \end{array}
  \right.
\end{equation}
where the velocity average is
\begin{equation}
\label{eq0364}
  (v_x)_{i',j',k} = \frac{1}{2}
  \left[  \left(v_x\right)_{i,j',k}
        + \left(v_x\right)_{i+1,j',k} \right]
  .
\end{equation}
Second order accuracy in space and time is achieved in the same
fashion as for the fluid updates: a first order predictor combined
with a second order corrector with a piece-wise linear TVD
approximation to the fluxes.
This ends the description of the $\left( b_y \right)_{i,j',k}$
component of the magnetic field.
The update of the $\left( b_z \right)_{i,j,k'}$ follows the same
strategy.

\subsection{Generalisation to non-uniform meshes}
\label{subsec:generalisation_to_non-uniform_meshes}
Finite volume methods can be constructed for non-uniform meshes in a
straightforward manner.
The only quantities affected by the non-uniform mesh are the volumes
of the computational cells, their bounding surfaces as well as the
cell edge lengths.

In the current version of the code, we have implemented irregular
Cartesian meshes.
Then the mesh increments $\Delta x_i=x_{i'}-x_{i'-1}$,
$\Delta y_j=y_{j'}-y_{j'-1}$, $\Delta z_k=z_{k'}-z_{k'-1}$
are no longer constant for the respective direction and equation
\eqref{eq0101} then changes to
\begin{equation}
\label{eq037}
    \frac{\partial \boldsymbol{u}_{i,j,k}}{\partial t}
  + \frac{1}{\Delta x_i} \left( \boldsymbol{F}_{i',j,k} - \boldsymbol{F}_{i'-1,j,k} \right)
  = 0
  .
\end{equation}
For more general coordinates see for example \cite{Vinokur1989}.

The update formulas for irregular Cartesian meshes are then simply
obtained by substituting $\Delta x$ by $\Delta x_i$ adequately in all
the previous sections.
Analogously $\Delta y$ by $\Delta y_j$ and $\Delta z$ by $\Delta z_k$.
Furthermore, the velocity interpolation in the magnetic field advection
needs also to conform with the non-equidistant spacing of the cell
centers.
However, the velocity averaging \eqref{eq0362} is left unchanged since
its primary goal is not an interpolation but an inclusion of the
stabilising effect of numerical dissipation.
Finally, the CFL condition eq. \eqref{eq0083} needs to be adapted,
\begin{equation}
\label{eq0420}
  \Delta t^n = k
      \min\left( \frac{\Delta x_i}{C_{i,j,k}^{n,x}}
                ,\frac{\Delta y_j}{C_{i,j,k}^{n,y}}
                ,\frac{\Delta z_k}{C_{i,j,k}^{n,z}} \right)
  \, \text{ for all } \, i,j,k
  .
\end{equation}

\section{Parallelisation and performance}
\label{sec:parallelisation_and_performance}
The implementation of the above-described simple algorithms uses the
directional operator splitting in a peculiar way: Instead of the traditional
approach to hold the data locations fixed in memory while sweeping updates
in $x$-, $y$-, and $z$-directions, we rearrange the data in the
memory between the sweeps so that different directional sweeps always
occur along the contiguous direction of the data in the memory
\cite{2003ApJS..149..447P}.
This has the advantage that the data load and store operations are
very efficient for the large arrays that contain the three-dimensional
data and that only one one-dimensional subroutine is required per
operator split physics ingredient to perform the corresponding data
update in the sweep. The disadvantage is the additional compute load
to rearrange the data (of order 10\% of the total CPU-time) and the
complications the rearranged data can cause if the code needs to import
or export oriented data between the sweeps (e.g. for debugging).

The most convenient operation to rearrange the data in the desired
way is a rotation with angle $2\pi/3$ about the axis threading the
origin and point $\left(1,1,1\right)$. Each rotation aligns another
original coordinate axis with the current x-direction, without changing
any relative quantities between data points or the parity of the system.
Three consequential rotations lead back to the original state. But
how does this procedure interfere with the parallelisation? After
having applied a cuboidal domain decomposition for distributed memory
architectures, one realises that it is not meaningful to rotate the
whole domain about the same rotation axis, because this would invoke
excessive data communication among the processes. It is sufficient
to to rotate all cuboids individually about the axis defined in their
local reference frame. These local rotations do not require any data
communication.

However, data communication is required across the interfaces of the
distributed data to calculate the advection terms in Eq. \eqref{eq0101}, the
gradient of the gravitational potential \eqref{eq02513}, or to interpolate the
velocities in the update of the magnetic field to the zone edges in
Eq. \eqref{eq0361} and \eqref{eq0362}. In the following, we describe our
approach to the parallelisation in two dimensions, its extension to the third
dimension is straightforward.

\begin{figure}
\centering
\includegraphics[angle=-90,width=6cm]{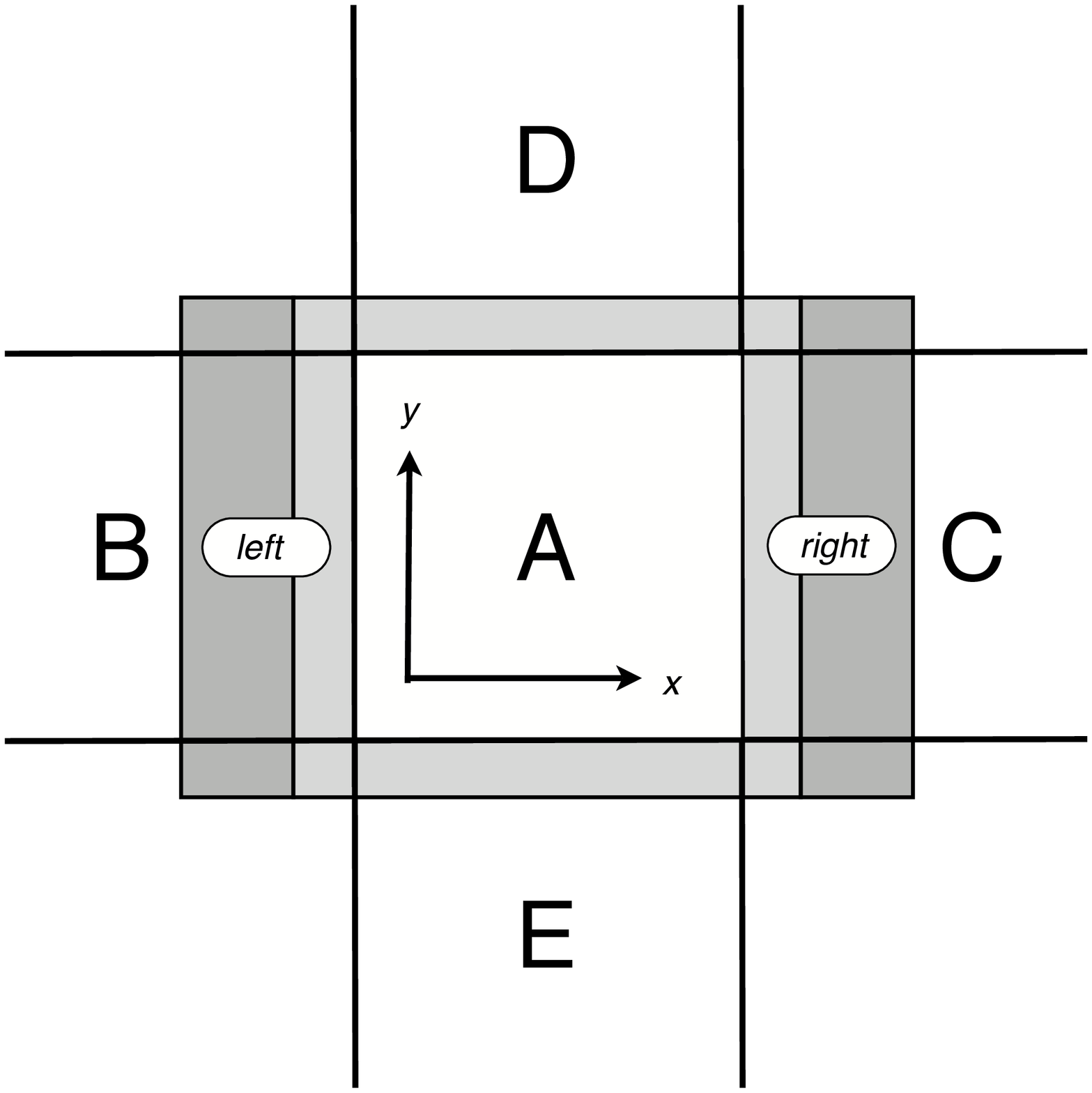}
\includegraphics[angle=-90,width=6cm]{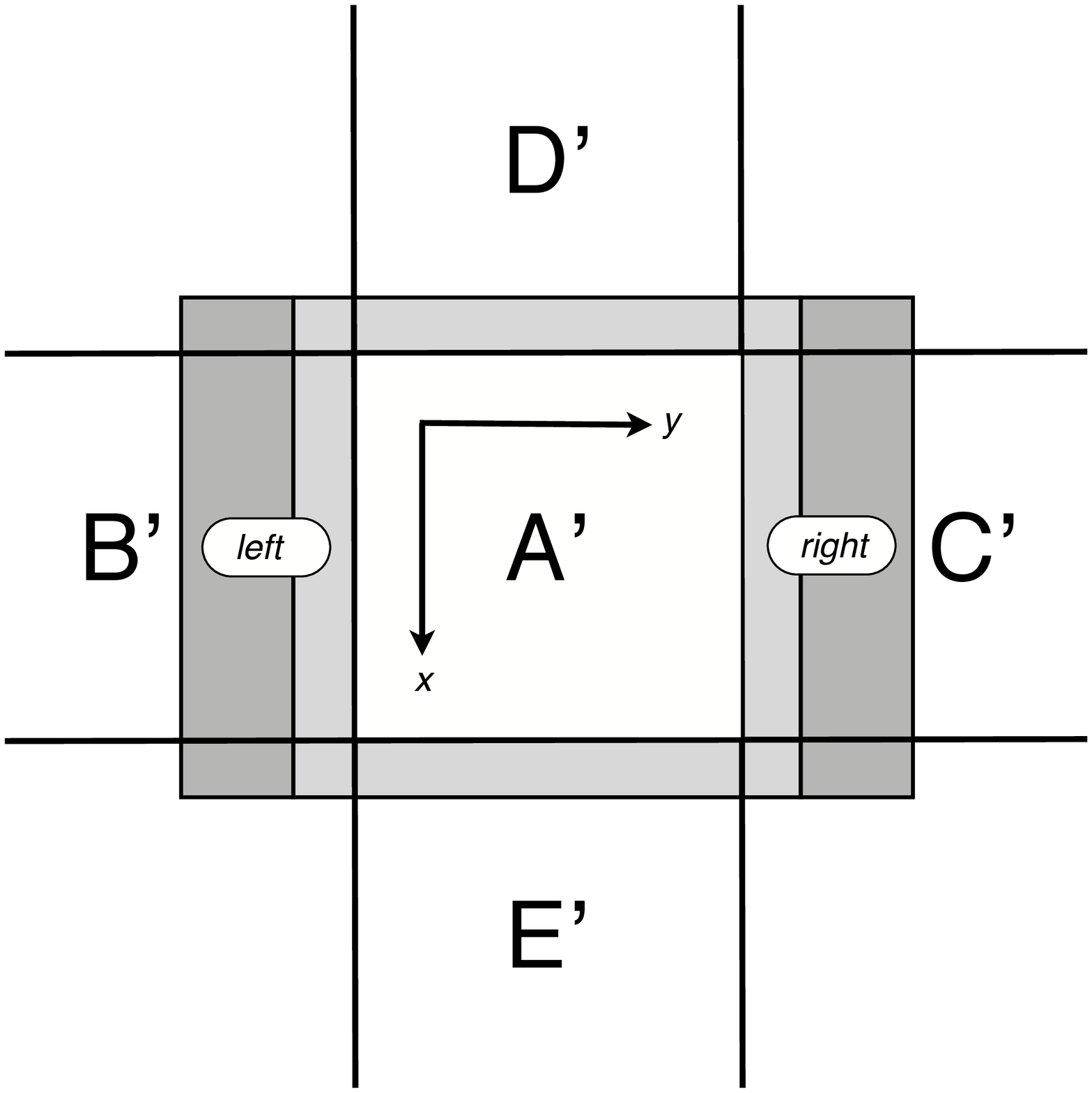}
\caption{The interaction between sweeps and rotations is illustrated for a
process A with its four neighbouring processes B-E. The distributed
data stored and treated in process A carries a small permanent buffer
area (light shading) around all of its interfaces. An additional volatile
buffer (dark shading) is added in the current $x$-direction. The
$x$-sweep is performed horizontally. For the $y$-sweep, the data
and the permanent buffer are rotated clockwise by $90$ degrees such
that the $y$-sweep can also be performed in the horizontal direction. }

\label{Flo:parallel.ps}
\end{figure}
Figure \ref{Flo:parallel.ps} shows the computational domain of a
process A after the cuboidal domain decomposition. The data stored
and updated in process A is surrounded by a small permanent buffer
zone (light shading). In the current $x$-direction, there is an additional
volatile buffer zone (dark shading). At the beginning of a time step,
the overlapping data from process B is communicated to process A in
order to update the buffer designated by 'left'. The overlapping data
communicated from process C fills the buffer designated by 'right'.
During the communication, the horizontal sweep in $x$-direction can
already start to work on the interior zones that don't require the
communicated buffer zones. Once all data has arrived, the $x$-sweep
can be completed so that all zones in domain A and the permanent buffer
is up to date.
Hence communication is overlapped with computation.

For the sweep in $y$-direction, all local data in Figure \ref{Flo:parallel.ps}
are rotated clockwise by $90$ degrees so that the $y$-direction
becomes horizontal. This time it is the data of process E that need
to be communicated to the buffer zones designated by 'left' and the
data of process D that need to be communicated to the buffer zones
designated by 'right'. Again, the sweep can start with the inner zones
and update the border zones once the communications have completed.
The important point is to realise that the $y$-sweep is now also
applied in the horizontal direction so that the one-dimensional subroutines
used for the $x$-sweep can be used without any modifications. In
the three-dimensional code, the procedure is repeated a third time
for the sweeps in $z$-direction. Finally, all three sweeps are once
more applied in reverse order to obtain second order accuracy. This
ordering of the sweeps requires four rotations between the six directional
sweeps.

Hence, the communication pattern is the only action that needs to
be performed differently during the different sweeps: For the $x$-sweep
we need to copy $B\longrightarrow A\left(\textrm{left}\right)$ and
$C\longrightarrow A\left(\textrm{right }\right)$, while for the $y$-sweep
we need to copy $E\longrightarrow A\left(\textrm{left}\right)$ and
$D\longrightarrow A\left(\textrm{right}\right)$. This distinction
can easily be implemented using persistent communications in the MPI.
When the code is initialised, the communications $B\longrightarrow A\left(\textrm{left}\right)$
and $C\longrightarrow A\left(\textrm{right }\right)$are defined to
be invoked by a handle $h_{x}$, while the communications $E\longrightarrow A\left(\textrm{left}\right)$
and $D\longrightarrow A\left(\textrm{right}\right)$are defined to
be invoked by a handle $h_{y}$. With this preparation it is straightforward
to perform the $x$-sweep by calling a function $\mathtt{sweep}\left(h_{x},n_{x},n_{y},A,\mathtt{f}\right)$
and the $y$-sweep by calling the same function $\mathtt{sweep}\left(h_{y},n_{y},n_{x},A',\mathtt{f}\right)$,
where $n_{x}$ and $n_{y}$ specify the dimensions of $A$, where
$A'$ is the rotated state of $A$ and where $\mathtt{f}$ points
to a one-dimensional function that implements the physical equations
treated by the sweep. In this way, the setup of the communication
topology during the initialisation of the code, the trigger of communications
and the dimensional index gymnastics in the $\mathtt{sweep}$ subroutine,
and the implementation of the physics in the function $\mathtt{f}$
are all very well disentangled.

The repeated evaluation of the physics equations in the function $\mathtt{f}$
amounts to the dominant contribution to the total CPU-time. Because
the sweeps are now always performed along contiguous memory, it is possible
to pipeline the physics quantities in the cache so that the access
of the large data arrays is reduced to a minimum. The first order
predictor and second order corrector are evaluated according to the
following scheme:

\begin{verbatim}
loop over cells i in x-direction
  u3 = u4
  u4 = u5
  u5 = u6
  u6 = u(i)
  if i<3 cycle
  uu1 = uu2
  uu2 = uu3
  uu3 = uu4
  uu4 = uu5
  uu5 = u5 + rate(u4,u5,u6)*0.5*dt !first order
  if i<7 cycle
  u(i-3) = u3 + rate(uu1,uu2,uu3,uu4,uu5)*dt !second order
end loop over cells
\end{verbatim}
Here, $u(i)$ is the state vector with the conserved variables, which
is only involved twice per time step $dt$, once for data retrieval
and once for data storage. The hydrodynamics equations of Eq. \eqref{eq0251}
and their discretisation described in Eqs. \eqref{eq0253} and \eqref{eq0255} are
abbreviated here by the evaluation of $\mathtt{rate()}$. The whole
operation has a stencil of $7$ cells and will lead to $3$ unassigned
cells at each end of the array $u$. Hence, the buffer in Fig. \ref{Flo:parallel.ps}
has to be large enough to host the unassigned cells. 

Note that the horizontal sweep in $x$-direction is performed independently
for all $n_{y}$ rows of the data in process $A$. Hence it is straightforward
to further parallelise the loop over the rows $i_{y}=1\ldots n_{l}$
with OpenMP, where $n_{l}=n_{y}+2n_{bp}$ is the dimension of $A$
including the permanent buffer of width $n_{bp}$ on either side.
This one OpenMP parallel section suffices to parallelise over 90\%
of the workload of the code for shared memory nodes. Hence, the above-described
approach naturally leads to a hybrid parallelisation, where MPI is
used to distribute the memory by the cuboidal domain decomposition
across different nodes or processors, while OpenMP is used to parallelise
the loop over the rows along the current sweep direction on the cores
that are available to each node or processor.

We evaluated the strong scaling of \noun{fish} on the new ROSA system (Cray XT-5,
nodes with 2 quad-core AMD Opteron 2.4 GHz Shanghai processors, SeaStar
2.2 communications processor with 2 GBytes/s of injection bandwidth
per node) at the Swiss National Supercomputer Center (CSCS). The dominant
limitation to the scaling of \noun{fish} is the rather large stencil that
emerges from the combination of the MHD and hydrodynamics solver with
a first and second order step in a single sweep. \noun{fish} scales without
problem if the problem size is increased with the increased number
of processors. More interesting is the case of a fixed size problem.
Figure \ref{fig:scaling.eps} shows the strong scaling for a problem with
$600^{3}$ cells. As the number of processors is increased, the ratio
of buffer zones to volume zones increases as well. In this case it
is the evaluation of the physics equations on the buffer zones that
limit the efficiency. However, Fig. \ref{fig:scaling.eps} also shows
that \noun{fish} scales very nicely to of order $10000$ processors in the
hybrid MPI/OpenMP mode where MPI is used between nodes and OpenMP
within the node. This scaling can be achieved because the parallelisation
with OpenMP does not increase the number of buffer zones. A version
of \noun{fish} with smaller stencils is under development.
\begin{figure}
\centering
\includegraphics[width=7cm]{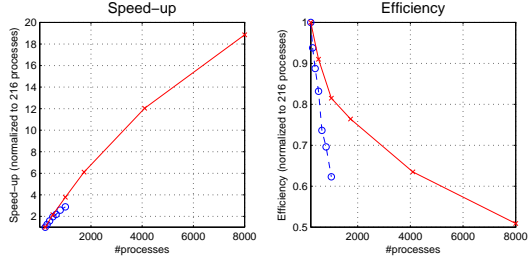}
\caption{Strong scaling of the \noun{fish} code. The speedup, normalised to 216 processes,
is shown on the left hand side. The efficiency is displayed on the
right hand side. The dashed lines with data points taken at the circles
refer to a parallelisation that uses only MPI. The solid lines with
data points taken at the crosses refer to a hybrid parallelisation
with MPI between nodes and OpenMP within nodes. The problem size was
kept constant at $600^{3}$ cells. The deviation from perfect scaling
is rather due to the increase of work on buffer zones with respect
to the work on volume zones than a bottleneck in the communication.
\label{fig:scaling.eps}}
\end{figure}

\section{Numerical results}
\label{sec:numerical_results}
In this section we verify our code by performing several
multidimensional test simulations of astrophysical interest.
The algorithm has already been tested for second order accuracy in
\cite{2003ApJS..149..447P} and we do not repeat that here.
Unless otherwise stated, we use periodic boundary conditions for all
test problems.
The simulations are stopped before any interaction due to the periodic
boundary can occur.

\subsection{Circularly polarised Alfv\'en waves}
\label{subsec:circularly_polarised_alfven_waves}
Our first test problem involves the propagation of circularly polarised
Alf\'en waves.
These waves are a smooth exact non-linear solution to the equations of
ideal MHD \cite{Landau1991a} and represent therefore an ideal problem
to test the convergence of numerical methods.

We have used the same initial conditions \cite{2000JCoPh.161..605T}.
Circularly polarised Alfv\'en waves are propagated the $x-y$ plane at
an angle of $\alpha=30^\circ$ with respect to the $x$-axis.
The domain is set to $[0,1/\cos\alpha]\times[0,1/\sin\alpha]$ and
discretised with $N\times N$ cells where $N=32,64,128,256,512$.
The adiabatic index is set to $\gamma=5/3$.
Then we have $\rho=1$, $v_\parallel=0$, $p=0.1$, $b_\parallel=1$,
$v_\perp =b_\perp  = A \sin[2\pi(x \cos\alpha + y \sin\alpha)]$ and
$v_z = b_z = A \cos[2\pi(x \cos\alpha + y \sin\alpha)]$.
The subscript $\parallel$ and $\perp$ denote the components parallel
and perpendicular to the wave propagation, respectively.
We have set the wave amplitude to $A=0.1$.
With these parameters the Alfv\'en velocity is $B_\parallel\rho=1$
and therefore the wave profile is advected to its initial position
after a simulation time $t=1$.

In figure \ref{fig0005} are shown the numerical errors after the
wave profile crossed the domain once $t=1$.
The error is given by the L1 norm of the difference between the
initial conditions and the numerical solution at time $t=1$ over
all cells:
\begin{equation}
  \delta \boldsymbol{u} = \frac{1}{N^2}\sum_{i,j} \mid \boldsymbol{u}_{i,j}^n - \boldsymbol{u}_{i,j}^0 \mid \nonumber
  .
\end{equation}
To get the absolute numerical error we have summed over all components
of $\delta \boldsymbol{u}$.
The scheme shows second order accuracy as can be inferred from
figure \ref{fig0005}.

\begin{figure}[!bh]
  \centering
  \includegraphics[scale=0.35]{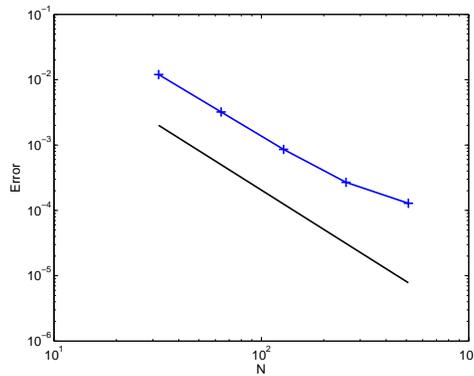}
  \caption{Convergence for circularly polarised Alfv\'en waves.
           The points are the absolute numerical error as function
           of the number of grid points.
           The solid line is a reference second order rate.
           }
  \label{fig0005}
\end{figure}

\subsection{Sedov-Taylor blast wave}
\label{subsec:sedov-taylor_blast_wave}
The Sedov-Taylor blastwave test is a purely hydrodynamical test
involving a strong spherically symmetric outward propagating shock wave.
An analytical self-similar solution can be found for example in
\cite{Landau1991,Whitham1974}.
We use an adiabatic index of $\gamma=5/3$.
The problem is solved on a cubic domain $(x,y,z)\in[0,1]^3$ with
$256^3$ computational cells and equidistant mesh spacing.

The density is set to unity and the velocity to zero throughout the
whole domain.
A huge amount of internal energy $e=3\times10^3$ is placed uniformly in a
small spherical region of radius $0.01$ at the center of the domain.
Outside the spherical region the energy is set to
$e=1\times10^{-3}$.
The energy amount concentrated at the center then starts a strong
shock	wave which propagates spherically outward from the center
of the domain.
The simulation was stopped at time $t=1.7419 \times 10^{-4}$ when the
shock has propagated to a distance $r\approx0.45$ from the center.
In figure \ref{fig001} a random subset of cells (blue points) and a
radial average of the numerical solution (green) are compared against the
exact solution (red).
All the values have been normalized to the exact postshock value.
The postshock values of the numerical solution are lower but come
close to the exact solution.
Despite we have a dimensionally split code, the scattering of the points
at the shock is not dramatic when compared to the Cartesian grid size of
$\Delta x = \Delta y = \Delta z \approx0.004$.
In figure \ref{fig002} a shaded surface plot is shown for the density
in the $x,y$ plane with $z=1/2$.
As seen in the figure, the shock looks spherical and no serious
symmetry breaking due to the dimensionally split character of the code
can be seen.

\begin{figure}[!bh]
  \centering
  \includegraphics[scale=0.35]{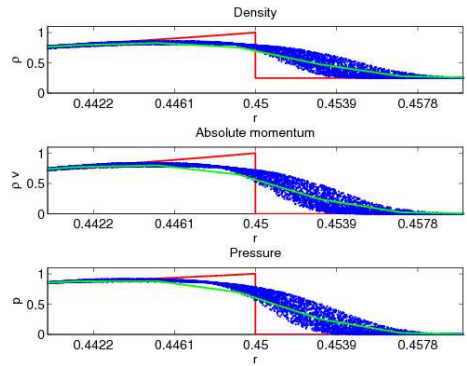}
  \caption{Radial distribution of the Sedov-Taylor blast wave test.
           The blue points represent a random subset of all the cells,
           the green line is a radial average of the numerical solution and the
           red line is the exact solution.
           All values have been normalized to the exact postshock values.
           The ticks on the $r$ axis have the size of the mesh spacing $\Delta x$
           and the shock is resolved within roughly $2\Delta x$.
           }
  \label{fig001}
\end{figure}

\begin{figure}[!bh]
  \centering
  \includegraphics[scale=0.35]{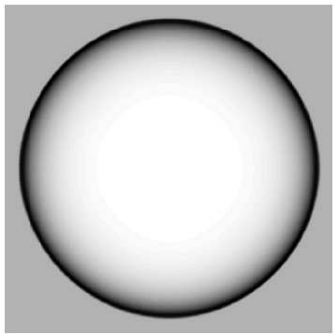}
  \caption{Shaded surface plot of the density for the Sedov-Taylor blast
           wave in the $x,y$ plane with $z=1/2$ illustrating the spherical
           symmetric character of the numerical solution.}
  \label{fig002}
\end{figure}

\subsection{Spherical Riemann problem}
\label{subsec:spherical_riemann_problem}
This problem is a spherical setup of Sod's shocktube and was suggested
in \cite{Toro1997} as a test for multidimensional hydrodynamics codes.
The solution is computed on a cubic domain $(x,y,z)\in[0,1]^3$ with
$256^3$ computational cells and equidistant mesh spacing.
Once again, an adiabatic index $\gamma=5/3$ is used.

The domain is decomposed into a spherical region of radius $R=0.25$
at the center of the domain and the remaining exterior region.
The fluid variables are set to constant values in each region.
At the boundary between both regions a circular discontinuity results.
In both regions, the velocity is set to zero.
In the inner region we set the density $\rho=0.125$ and the pressure
$p=0.1$.
In the outer region, we set the density $\rho=1$ and the pressure
$p=1$.
The simulation is stopped at time $t=0.09$.

In figure \ref{fig003} the density distribution at the end of the
simulation is displayed.
The blue points are a random subset of cells, the green line is a
radial average of the numerical solution and the red line is a highly
resolved numerical solution from a 1D spherically symmetric code based
on the same algorithm as the 3D one.
The shock, the contact discontinuity and the rarefaction wave are
well captured by the 3D code.
The scattering of the points is strongest at the inward propagating
shock.
In figure \ref{fig004} a shaded surface plot of the density
in the $x,y$ plane with $z=1/2$ is pictured.
The spherical character of the solution is well retained and
the dimensional splitting does not degrade this symmetry.

\begin{figure}[!bh]
  \centering
  \includegraphics[scale=0.35]{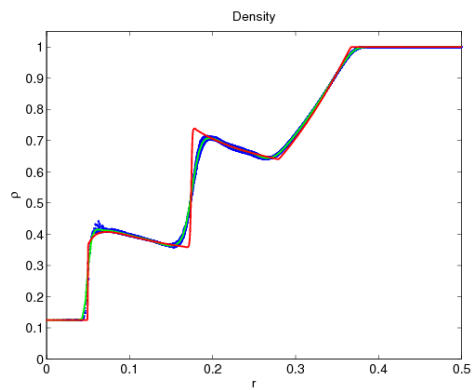}
  \caption{Radial density distribution of the spherical Riemann problem.
           The blue points represent a random subset of all the cells,
           the green line is an average over all the cells and the red
           line is the result of a highly resolved 1D simulation.}
  \label{fig003}
\end{figure}

\begin{figure}[!bh]
  \centering
  \includegraphics[scale=0.35]{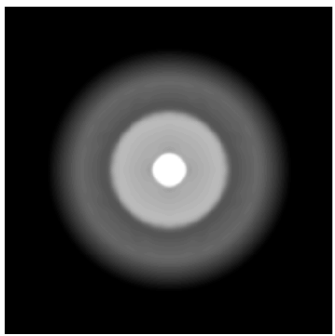}
  \caption{Shaded surface plot of the density for the spherical Riemann problem
           in the $x,y$ plane with $z=1/2$ illustrating the spherically symmetric
           character of the numerical solution. The density ranges from 1 (black)
           to 0.125 (white).}
  \label{fig004}
\end{figure}

\subsection{Magnetic explosion}
\label{subsec:magnetic_explosion}
This test is based on the same idea as the Sedov-Taylor blastwave
problem with the addition of a magnetic field
We used the same domain, number of cells and adiabatic index.
The density is set to $\rho=1$ and the velocity to vanish everywhere in
the domain.
At the center of the domain we set the pressure $P=100$ in a spherical
region with radius $R=0.1$.
Outside of a sphere with radius $R=0.125$ we set the pressure to
$P=1$.
In the spherical shell between the high central pressure and the outer
low pressure regions, we let the pressure vary linearly between the
two extremes.
The magnetic field components are set as $b_x=b_y=7/\sqrt(2)$
and $b_z=0$.
Inside the high pressure sphere the ratio between gas pressure and
magnetic pressure is $\beta=2P/B^2\approx4$ and outside it is
$\beta=2P/B^2\approx0.04$.
This is a difficult test problem for codes evolving the total
energy:
The internal energy is obtained by the substraction of kinetic and
magnetic energy from the total energy.
When the energy density is locally dominated by the magnetic field,
negative internal energies can occur which then break down the
simulation.

The simulation was stopped at $t=0.03$ and we used a CFL number
of $0.5$.
For the fluid update we used the minmod limiter.
The results are displayed in figure \ref{fig005} for the density, pressure,
kinetic energy and magnetic energy.
As apparent from the figure, spherical symmetry is broken and one
clearly distinguishes between the flow propagation parallel and
orthogonal to the magnetic field.
The outermost shell indicates a fast magnetosonic shock, which
is only weakly compressive.
The energy density is dominated by the magnetic field.
On the inside there are two dense shells propagating parallel to the
magnetic field.
From the outside these dense shells are bound by a slow magnetosonic
shock and a contact on the inside.

Similar initial conditions can be found in the literature, see e.g.
\cite{Balsara1999,2000ApJ...530..508L,zachary:263}.
We have tried to run our code with the same initial conditions as
in ref. \cite{Gardiner2008} where the magnetic field is higher
$b_x=b_y=10/\sqrt(2)$ and with no linear interpolation between
the high and low pressure regions.
However, we didn't succeed under reasonable CFL conditions
($>0.05$) and simulation time.
The appearance of negative internal energies always stopped
the simulation.
For very low $\beta$ flows we pay the price for the algorithmic
simplicity of splitting fluid and magnetic advection.
An upwind interpolation of the velocities to cell faces might help to
cure the problem.
However, we have not yet explored this possibility.

\begin{figure}[!bh]
  \centering
  \includegraphics[scale=0.6]{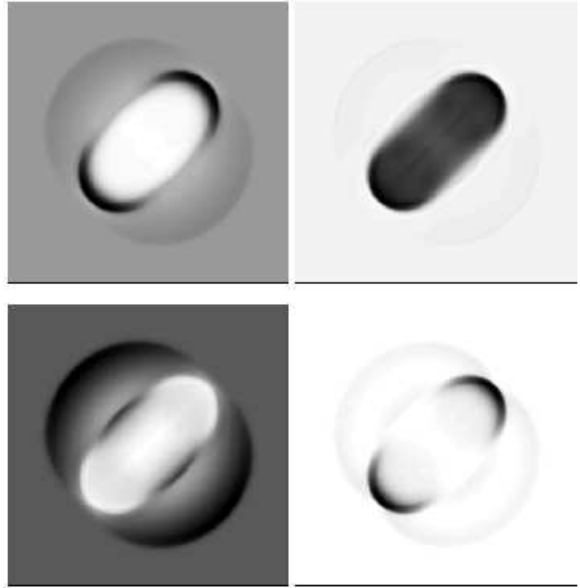}
  \caption{Shaded surface plot of the magnetic explosion at time $t=0.03$.
  Density (top left) ranges from 0.24 (white) to 2.17 (black).
  Gas pressure (top right) ranges from 0.25 (white) to 15.85 (black).
  Magnetic energy (bottom left) ranges from 7.66 (white) to 33.50 (black).
  Kinetic energy (bottom right) ranges from 0 (white) to 17.33 (black).}
  \label{fig005}
\end{figure}

\subsection{The rotor problem}
\label{subsec:the_rotor_problem}
Our next test is the so-called rotor problem, presented in
\cite{Balsara1999}.
We use the first rotor problem in \cite{2000JCoPh.161..605T} as initial
conditions.
We solve this 2D problem in the Cartesian domain $[0,1]^2$ with $256^2$
computational cells.
The problem consists of a dense rapidly rotating cylinder (the rotor)
with $\rho=10$, $v_x=-2(y-1/2)/r_0$, $v_y=2(x-1/2)/r_0$
extending up to a radius $r_0=0.1$, where $r=[(x-1/2)^2+(y-1/2)^2]^{1/2}$,
installed in a lighter resting fluid.
The lighter fluid is characterized by $\rho=1$, $v_x=v_y=0$ for
$r>r_1=0.115$.
In between the rotating and the light fluid $r_0<r<r_1$ we set
\begin{eqnarray}
  \rho & = & 1 + 9 f(r) \nonumber \\
  v_x & = & - 2 f(r) ( y - 1/2 ) / r \nonumber \\
  v_y & = &   2 f(r) (x - 1/2 ) / r \nonumber
\end{eqnarray}
where
\begin{equation}
  f(r) = \frac{r_1-r}{r_1-r_0}.
  \nonumber
\end{equation}
This smoothes out the discontinuities and reduces initial
transients.
The magnetic field is initially set to $b_x=5/\sqrt{4 \pi}$, $b_y=b_z=0$
and the pressure is uniformly set to $p=1$.
The adiabatic index used for this test is $\gamma=1.4$.

The simulation was run to time $t=0.15$ and the results are displayed
in figure \ref{fig006}.
The dense rotating cylinder initially not in equilibrium has started to
expand until the magnetic pressure due to field wrapping has stopped
the expansion resulting in its oblate shape.
The outer layers of the dense cylinder has lost part of its initial
angular momentum in form of Alv\'en waves radiating away.
This braking of the magnetic rotor is a possible model for the angular
momentum loss of collapsing gas clouds in star formation
\cite{1980ApJ...237..877M,1980M&P....22...31M}.

\begin{figure}[!bh]
  \centering
  \includegraphics[scale=0.6]{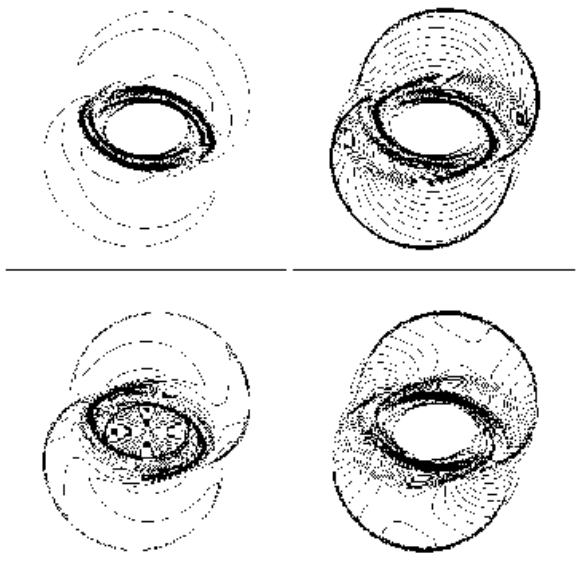}
  \caption{Contour plot of the rotor problem at time $t=0.15$.
  The density (top left),
  the gas pressure (top right),
  the Mach number (bottom left) and
  the magnetic pressure (bottom right).
  }
  \label{fig006}
\end{figure}

\subsection{Two dimensional MHD Riemann problem}
\label{subsec:two_dimensional_mhd_riemann_problem}
The next test we performed is a two dimensional Riemann problem.
The initial conditions are similar to \cite{Dai1998331}:
\begin{equation}
  \left( \rho,v_x,v_y,v_z,P \right)
  = \left\{
    \begin{array}{r r r r}
      \left( \right. 1 , & -0.75 ,&  0.5 ,& 0. , 1. \left. \right) , \; x \leq L_x/2 , \; y \leq L_y/2 \\
      \left( \right. 3 , & -0.75 ,& -0.5 ,& 0. , 1. \left. \right) , \; x > L_x/2    , \; y \leq L_y/2 \\
      \left( \right. 2 , &  0.75 ,&  0.5 ,& 0. , 1. \left. \right) , \; x \leq L_x/2 , \; y > L_y/2    \\
      \left( \right. 1 , &  0.75 ,& -0.5 ,& 0. , 1. \left. \right) , \; x > L_x/2    , \; y > L_y/2    \\
    \end{array}
  \right.
  \nonumber
\end{equation}
and the magnetic field is uniformly $\boldsymbol{b}=\left( 2,0,1 \right)/\sqrt{4\pi}$.
We setup the problem on the domain $[0,0.8]^2$ and hence $L_x=0.8$,
$L_y=0.8$ with $512^2$ cells.
The adiabatic index is set to $\gamma=5/3$.
For this test we used zeroth order extrapolation boundary conditions
and evolved the problem to time $t=0.8$
The density, pressure, kinetic energy and magnetic energy are displayed
in figure \ref{fig007}.
A similar test problem can also be found in \cite{Serna20094232}.
Our results compare qualitatively well with the cited references.

\begin{figure}[!bh]
  \centering
  \includegraphics[scale=0.6]{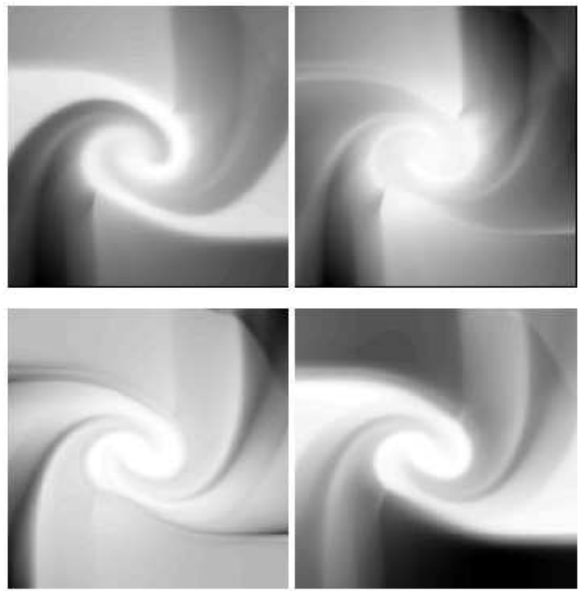}
  \caption{Contour plot of the 2D MHD Riemann problem at time $t=0.8$.
  Density (top left) ranges from 0.3428 (white) to 2.5698 (black).
  Gas pressure (top right) ranges from 0.1208 (white) to 0.7757 (black).
  Magnetic energy (bottom left) ranges from 0.0039 (white) to 0.2953 (black).
  Kinetic energy (bottom right) ranges from 0 (white) to 0.6233 (black).
  }
  \label{fig007}
\end{figure}

\subsection{Steady polytrope}
\label{subsec:steady_polytrope}
As a final test, we show the performance of our HSE correction scheme on
an astrophysical problem.
We simulate a model star, a so-called polytrope \cite{1967aits.book.....C}.
These model stars are constructed in spherical symmetry from hydrostatic
equilibrium
\begin{equation}
  \label{eq0430}
  \frac{\ud p}{\ud r} = - \rho \frac{\ud \phi}{\ud r}
\end{equation}
and Poisson's equation
\begin{equation}
  \label{eq0440}
  \frac{1}{r^2} \frac{\ud}{\ud r} \left( r^2 \frac{\ud \phi}{\ud r} \right) = 4 \pi G \rho
\end{equation}
where $r$ is the radial variable and $G$ is the gravitational constant.
The functional relation between $p$ and $\rho$ is of the form
$p=\kappa \rho^\gamma$ where $\kappa$ and $\gamma$ are constants.
This relation is called a polytropic equation of state, $\kappa$ is the
polytropic constant and $\gamma$ the polytropic exponent not to be
confused here with the adiabatic index.
We chose the following parameters: $M=1 M_\odot$,
$R=1 R_\odot$ and $\gamma=4/3$.
Here $M$ is the total mass of the star in units of solar masses and
$R$ is the radius of the star in units of solar radii.
With these parameters eq. \eqref{eq0430} and \eqref{eq0440} can easily
be solved by any ODE integrator.
The resulting star has then a polytropic constant
$\kappa\approx3.841\times10^{14}$ (in cgs units) and a mean density of
$\overline{\rho}\approx1.408$ $\mathrm{g/cm}^3$.
This sets the hydrostatic timescale
$\tau_{\mathrm{HSE}}$ $\approx$ $(\overline{\rho}G)^{-1/2}/2$ $\approx$ $1.631\times10^{3}\mathrm{s}$,
which is the characteristic timescale on which the star reacts to perturbations
of its hydrostatic equilibrium.
We then mapped the spherical profiles of the density, internal energy and
gravitational potential onto a 3D Cartesian grid of
dimensions $[0,2.9 R_\odot]^3$ such that the center of the star coincides
with the center of the domain.
We have put around the star a low density atmosphere $\rho_{\mathrm{atm}}=10^{-6} \rho_C$
where $\rho_C=76.313$ $\mathrm{g/cm}^3$ is the central density.
The low density atmosphere is not in hydrostatic equilibrium but since
its mass is so low it will have little dynamic effect on the star.
We adopted continuous boundary conditions.

We then evolve the hydrodynamic variables for $20$ hydrostatic
timescales $\tau_{\mathrm{HSE}}$ with the ideal gas law \eqref{eq0050},
$\gamma=4/3$, and with the unsplit source term integrator described in
\ref{subsec:incorparation_of_gravitation}.
The gravitational potential is kept fixed for the whole simulation.
We have used both $64^3$ and $128^3$ cells to discretise the domain.

The results of the simulation are shown in  figure \ref{fig008}.
The subplot on the left hand side shows the spherically averaged density
distribution of the simulation with $128^3$ cells resolution with and
without the HSE correction (dash dotted and solid red lines
respectively).
The solid blue line is the density profile as initially put on the 3D
grid and is therefore the reference solution.
The subplot on the right hand side shows the central density normalised
to the exact value as function of time in units of the hydrostatic
timescale $\tau_{\mathrm{HSE}}$.
The dash dotted lines are once again with the HSE correction while
the solid lines are without.
The blue lines have been computed with $64^3$ cells and the red lines
with $128^3$ cells.

Due to the fact that the center of the star is an extremum in density
and internal energy, any TVD scheme switches there to first order
accuracy.
Hence the strong numerical dissipation associated with first order
accuracy heats the center of the star which increases the pressure and
as a consequence the central density is diminished.
In other words the star evaporates.

As can be seen from figure \ref{fig008}, the HSE correction
effectively lowers this spurious effect.
The HSE corrected simulation with $64^3$ cells (blue dash dotted on
right subplot) does even better than the simulation
with $128^3$ cells without correction.

With increasing resolution, as one expects, the effect of the HSE
correction gets less dramatic.
However, for low to intermediately resolved strong gravitationally
induced gradients the HSE correction can greatly improve the accuracy
at the center of the star.

\begin{figure}[!bh]
  \centering
  \includegraphics[scale=0.6]{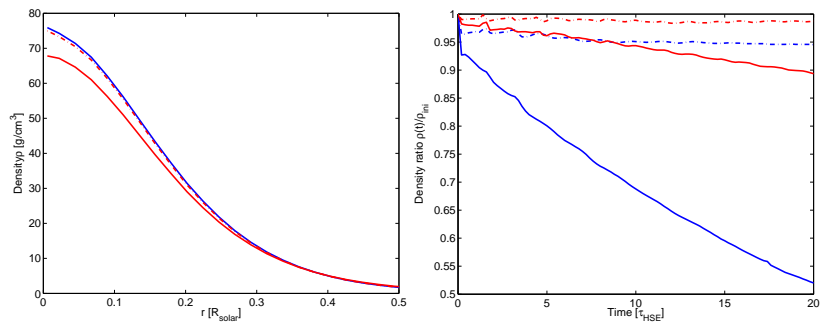}
  \caption{Left: radially averaged density for the simulation with $128^3$
                 cells with (dash dotted red line) and without
                 (solid red line) HSE correction.
                 The solid blue line is the reference solution.
                 The HSE correction significantly increases the accuracy in
                 the inner region.
           Right: central density normalised to the exact value as a function
                  of time. The dash dotted and the solid lines again denote the
                  simulation results with and without HSE correction.
                  The blue lines were computed with $64^3$ cells and the red ones
                  with $128^3$.
                  The central density is kept nearly at constant
                  value for the HSE corrected simulations while without the
                  correction the central density decreases monotonically.
                  Note the initial decrease in central density for all curves which is
                  an initial relaxation of the numerical solution to the employed
                  resolution.
  }
  \label{fig008}
\end{figure}

\section{Conclusion}
\label{sec:conclusion}
We describe a simple and efficient three-dimensional implementation of the ideal
magneto-hydrodynamics equations in our code named \noun{fish}. The algorithms used are mostly
based on \cite{2003ApJS..149..447P}. Three different choices make the code very transparent and
easy to work with: First, the rigorous operator splitting in second order operator
ordering separates the sweeps across different coordinate directions. Also the
constrained transport of the magnetic field is operator split from the hydrodynamics
update. Second, the use of a Riemann solver free relaxation scheme with second order TVD advection facilitates the update of the state vector. Second order in time is
achieved by a predictor-corrector approach within each time step. Third, subroutines for
the physics operators are only required for the $x$-direction because the sweeps in
$y$-direction and $z$-direction are performed by a rotation of the computational domain.
Between the sweeps, the data is rotated such that an $x$-sweep in the direction of
contiguous memory actually performs a $y$- or $z$-sweep in the coordinate frame of the
rotated data. We document the details of an improved discretisation of the advective
fluxes with respect to earlier versions of the code. Due to its clarity, we have
successfully used the code for exercises and computational experiments in
magneto-hydrodynamics classes.

We present different tests to evaluate the accuracy of \noun{fish}. Second order convergence is
demonstrated by the example of a circularly polarised propagating Alfven wave. The
capability of accurately capturing hydrodynamic shocks is shown in a Sedov-Taylor blast wave
explosion and in a spherical Riemann problem that are
evolved in the three-dimensional computational domain.
The latter two examples produce nice spherical flow features and show that the directional operator
splitting does not cause sizeable asymmetries.
We also investigate strong magnetosonic shocks in the form of a magnetic explosion.
For this problem we reached a limit of the current scheme for very low plasma $\beta$ and
very strong shocks.
However, there is still room for improving the scheme in this extreme regimes and future
developments will go in that direction.
In order to further test the coupling
between the magnetic fields and the fluid we perform the rotor problem in a
two-dimensional computational domain. Finally, we calculate a 2D MHD Riemann problem and
successfully compare the result with the literature. The only test that led to
insufficient performance was the time evolution of a gravitationally bound hydrostatic
object at low resolution. The problem is that the presence of gravitation leads to
stationary density gradients that are mistaken as propagating wave gradients in the
scheme that tailors the numerical dissipation for a total variation diminishing solution.
The corresponding stationary dissipation leads to the evaporation of the bound object on
few dynamical time scales. We remedied this problem by analytically estimating the
hydrostatic gradients and subtracting them from the total gradients before calculating
and applying the dissipation for the propagating waves. This to our knowledge novel and
simple measure could also be used in combination with the Roe Riemann solver and
potentially other approximate solvers. It allows the stable evolution of a
gravitationally bound object over many dynamical time scales without exaggerated
numerical dissipation at the center. This is shown in our last test case where we evolve
a polytrope in hydrostatic equilibrium.

Inspite of the straightforward approach, \noun{fish} is very efficient, even for problems
containing $\sim 1000^{3}$ and more cells. \noun{fish} implements cubic domain decomposition for
distributed memory computer clusters. The process-wise rotation of the data between the
sweeps takes about $10\%$ of the computational cost, but is largely compensated for by
the fact that the individual sweeps along contiguous memory can be pipelined in the
cache. The rotation of the data before the directional sweeps enables a very elegant
parallelisation scheme that is based on non-blocking persistent MPI communication
directives. We perform the communication for each direction separately and overlap it
with computations of the update of interior cells. As sweeps along the same direction are
embarassingly parallel within a shared memory cluster, we additionally parallelise those
using OpenMP directives. With this, \noun{fish} can take advantage of multi-core processors and
scales nicely to $\sim 10000$ processes for a problem of fixed size. The limiting factor
in the scaling of \noun{fish} is the increasing ratio of work on buffer zones to work on
enclosed zones. Even more processes can be used efficiently if the problem size grows
with the number of available processes.

\noun{fish} is mainly targeted for astrophysical applications where simple and efficient
low-order algorithms are appreciated to accept manifold variations of input physics. The
input physics is often a patchwork of different descriptions and algorithms for the
different domains. As they do not always connect with a prescribed smoothness and as
decisions about approximations are often taken during runtime, it is difficult to work
with higher order methods. \noun{fish} has successfully been used in one of the first
predictions of the graviational wave signal from 3D supernova models with microscopic
input physics \cite{2008A&A...490..231S} and provides the foundation for the implementation of spectral neutrino
transport in our new code ELEPHANT (ELegant and Efficient Parallel Hydrodynamics with
Approximate Neutrino Transport) \cite{2009ApJ...698.1174L}.
The dynamical contrast of the density in many astrophysical
applications suggests the implementation of an adaptive mesh. We discussed the extension
of the approach to non-uniform orthogonal meshes, but find them only applicable for
limited dynamical contrasts of about a factor of two increase of the resolution. On the
long term, it appears more promising to embed the uniform Cartesian mesh of our current
approach in a multi-patch driver like e.g. carpet in the cactus framework. We hope that
this paper will serve as a reference for future users and developers of the public
version of \noun{fish}, which is in preparation and will soon become available.

\section*{Acknowledgments}
We would like to acknowledge Hugh Merz for helping with the optimisation of the code
and Stephen Green for his explorations of irregular meshes.
This research was funded by the Swiss National Science Foundation under
grant SNF 200020-122287 and SNF PP0022-106627.
The scaling properties of the code have been investigated at the Swiss National
Supercomputing Centre CSCS.


\bibliographystyle{elsarticle-num.bst}


\let\jnl=\rm
\def\aap{\jnl{A\&A}}              
\def\apj{\jnl{ApJ}}                 
\def\apjl{\jnl{ApJ}}                
\def\apjs{\jnl{ApJS}}             


\end{document}